\documentclass[%
 reprint,
 amsmath,amssymb,
 aps,
 prb,
floatfix,
]{revtex4-2}

\usepackage[utf8]{inputenc}
\usepackage{graphicx}
\usepackage{dcolumn}
\usepackage{bm}
\usepackage{gensymb}
\usepackage{afterpage}
\usepackage[caption=false]{subfig}
\usepackage[colorlinks=false,pdfborder={0 0 0}]{hyperref}
\usepackage{cleveref}
\usepackage{siunitx}
\usepackage{color}

\graphicspath{ {figures/} }

\crefname{figure}{Fig.}{Figs.}
\crefname{equation}{Eq.}{Eqs.}

\let\origcite\cite
\def\cite#1{\unskip~\origcite{#1}}
\let\origcitep\citep
\def\citep#1{\unskip~\origcitep{#1}}

\makeatletter
\AtBeginDocument{
  \hypersetup{
    pdftitle = {\@title},
    pdfauthor = {\@author}
  }
}
\makeatother

\let\vec\mathbf

\begin{document}
\title{flatspin: A Large-Scale Artificial Spin Ice Simulator}
\author{Johannes H. Jensen}
\email{johannes.jensen@ntnu.no}
\author{Anders Strømberg}
\author{Odd Rune Lykkebø}
\author{Arthur Penty}
\author{Magnus Sj\"alander}
\author{Erik Folven}
\author{Gunnar Tufte}
\affiliation{Norwegian University of Science and Technology, Trondheim, Norway}

\date{\today}

\begin{abstract}
We present flatspin, a novel simulator for systems of interacting mesoscopic spins on a lattice, also known as artificial spin ice (ASI).
Our magnetic switching criteria enables ASI dynamics to be captured in a dipole model.
Through GPU acceleration, flatspin can simulate realistic dynamics of millions of magnets within practical time frames.
We demonstrate flatspin's versatility through the reproduction of a diverse set of established experimental results from the literature.
In particular, magnetization details of ``pinwheel'' ASI during field-driven reversal have been reproduced, for the first time, by a dipole model.
The simulation framework enables quick exploration and investigation of new ASI geometries and properties at unprecedented speeds.
\end{abstract}

\maketitle

\section{Introduction}
An artificial spin ice (ASI) is an ensemble of nanomagnets arranged on a lattice, coupled through magnetic dipole-dipole interactions.
The vast variety of emergent collective behaviors found in these systems have generated considerable research interest over the last decade \citep{skjaervo_advances_2020, Heyderman2013}.
Using modern nanofabrication techniques, emergent phenomena can be facilitated through direct control of the ASI geometry, e.g., collective ferromagnetic/antiferromagnetic ordering \citep{sklenar2019}, Dirac strings \citep{morris_dirac_2009}, and phase transitions \citep{levis_thermal_2013, anghinolfi2015}. 
ASIs offer a unique model system for exploring fundamental physics, since magnetic microscopy enables direct observation of their internal state.
There is also a growing interest in ASIs as building blocks for novel devices \citep{jensen2018}.

Micromagnetic simulations of ASI have been limited to a handful of nanomagnets due to excessive computational cost.
Although physically accurate, such high fidelity simulations are unable to capture large-scale emergent phenomena, such as the size of magnetically ordered domains and long-range order.
To simulate large ASI systems, an established approach is to sacrifice fidelity for speed by employing a dipole model, i.e., treating each nanomagnet as a single macro spin approximated by a point dipole \cite{budrikis_phd}.
Traditionally, Monte Carlo methods have been used in conjunction with the dipole approximation to search for low energy configurations \citep{Zhang2013, ke_energy_2008} or study statistical measures such as vertex populations \cite{budrikis_phd}.
However, Monte Carlo methods are inherently stochastic and better suited for ensemble statistics rather than detailed dynamics \citep{baryam}.

flatspin is a simulator for large ASI systems based on a dipole approximation with the ability to capture realistic dynamics, inspired by the work of \textcite{budrikis_phd}.
The flexibility of flatspin enables quick exploration of ASI parameters and geometries.
Through GPU acceleration, flatspin can capture realistic dynamics of millions of magnets within practical time frames.

In this paper, we present the motivation and design of flatspin.
We demonstrate good agreement between flatspin and a variety of published experimental results.
We show that flatspin can capture dynamic behaviors observed experimentally, which have previously eluded modeling\citep{Li2019}.

\section{\label{sec:modeling}The flatspin magnetic model}
In this section, we describe the dipole model and the underlying physical assumptions of flatspin.
The model is designed to simulate the ensemble state-by-state evolution, i.e., \textit{dynamics}, of two-dimensional ASI.
In short, magnets are modeled as point dipoles (\cref{sec:magnet_dipole}), and each dipole can be affected by three types of external influence: 
magnetic dipole-dipole coupling (\cref{sec:spin-spin}),
an applied external magnetic field (\cref{sec:external-field}),
and thermal fluctuations (\cref{sec:thermal-field}).
The switching of spins is determined using a generalized Stoner-Wohlfarth model (\cref{sec:switching}).
Imperfections in the ASI are introduced as different coercive fields, set per spin (\cref{sec:disorder}).
Dynamics are modeled using a deterministic single spin flip strategy (\cref{sec:dynamics}).

\subsection{\label{sec:magnet_dipole}Magnets as dipoles}

ASI systems are physically realized as elongated islands of a ferromagnetic material, arranged on a two-dimensional lattice.
The magnets are made small enough to exhibit a single ferromagnetic domain, i.e., coherent magnetization throughout the magnet.
The single domain is stable as the energy cost associated with domain walls exceeds the cost associated with the demagnetization energy\cite{frenkel-singledomain, kittel-singledomain}.
Since a magnet has coherent magnetization, it can be approximated by a single mesoscopic spin and the magnetic state can be represented by a single vector $\vec{m}$.

The magnets will exhibit an in-plane shape anisotropy due to their small thickness, as well as a magnetization direction defined by their elongated shape.
Hence individual magnets can be approximated by classical macro spins with a twofold degenerate ground state defined by the elongated shape of the individual elements.
Due to the two degenerate ground state configurations, we approximate each magnet as a magnetic dipole with \emph{binary} magnetization, i.e., a macro spin, $s_i \in \{-1, +1\}$.

\begin{figure}[t]
    \centering
    \def\svgwidth{0.9\linewidth}
    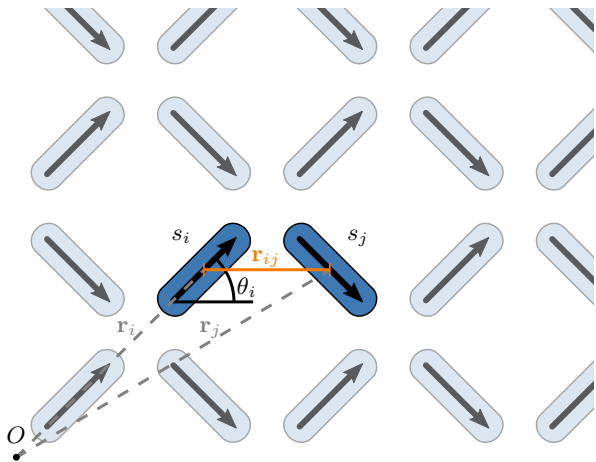
    \caption{\label{fig:spins}%
    The representation of nanomagnets as spins $s_i$ and associated quantities: angle $\theta_i$ and distance to neighbor $j$, $\vec{r}_{ij}$. Note that the magnetization of spin $i$ is given by its spin, $s_i$, and orientation, $\theta_i$. 
    }
\end{figure}

As illustrated in \cref{fig:spins}, each magnetic dipole is modelled with a position $\vec{r}_i$ and rotation $\theta_i$, which together define the ASI geometry.
Furthermore, each magnet is assigned a coercive field, $h_{c}^{(i)}$, describing its resistance to switching (see \cref{sec:switching}).
Using reduced units, the magnetization vector of a single magnet can be expressed as
\begin{equation}
    \vec{m}_i=s_i \vec{\hat{m}}_i
\end{equation}
where $\vec{\hat{m}}_i$ is the unit vector along $\vec{m}_i$.

\subsection{\label{sec:total-field}Magnetic Fields and Temperature}

External fields and temperature are modeled as a combination of effective magnetic fields.
The total field, $\vec{h}_i$, affecting each magnet $i$ is the sum of three components: 

\begin{equation}
    \vec{h}_i = \vec{h}_\text{dip}^{(i)} + \vec{h}_\text{ext}^{(i)} + \vec{h}_\text{th}^{(i)},
\end{equation}
where $\vec{h}_\text{dip}^{(i)}$ is the local magnetic field from neighboring magnets (magnetic dipole-dipole interactions), $\vec{h}_\text{ext}^{(i)}$ is a global or local external field, and $\vec{h}_\text{th}^{(i)}$ is a random magnetic field representing thermal fluctuations in each magnetic element.
The contributions from each field is described in the following sections. 

\subsection{\label{sec:spin-spin}Magnetic dipole-dipole interactions}

The individual magnets, or spins, are coupled solely through dipole-dipole interactions.
Each spin, $i$, is subject to a magnetic field from all neighboring spins, $j\neq i$, given by 
\begin{equation}
    \vec{h}_\text{dip}^{(i)} = \alpha \sum_{j \ne i}\frac{3\vec{r}_{ij}(\vec{m}_j \cdot \vec{r}_{ij})}{|\vec{r}_{ij}|^5} - \frac{\vec{m}_j}{|\vec{r}_{ij}|^3},
\end{equation}
where $\vec{r}_{ij}=\vec{r}_i-\vec{r}_j$ is the distance vector from spin $i$ to $j$, and $\alpha$ scales the dipolar coupling strength between spins.
The coupling strength $\alpha$ is given by $\alpha = \frac{\mu_0 M}{4\pi a^3}$, where $a$ is the lattice spacing, $M$ is the absolute magnetic moment of a single magnet, and $\mu_0$ is the vacuum permeability.
The distance $\vec{r}_{ij}$ is given in reduced units of the lattice spacing.

The dipole field present at each spin's location is calculated by summing the dipole field contributions from spins in its neighborhood. The size of the neighborhood is user-configurable and defined in units of the lattice spacing.
In some geometries, such as square ASI, short range interactions dominate the contributions to $\vec{h}_\text{dip}$\citep{Wang2006,chungpeng-ising-neighbors}, in which case the neighborhood size can be relatively small, for a benefit of increased efficiency.
For geometries where long range interactions are significant, a larger neighborhood is required, e.g., pinwheel ASI\citep{Macedo2018}.

\subsection{\label{sec:external-field}External field}

Applying an external magnetic field is the primary mechanism for altering the state of an ASI in a controlled manner.
The external field can either be set locally on a per-spin basis, $\vec{h}_\text{ext}^{(i)}$, globally for the entire system, $\vec{h}_\text{ext}$, or as a spatial vector field, $\vec{h}_\text{ext}(\vec{r})$.

Time-dependent external fields are supported, i.e., $\vec{h}_\text{ext}$ is a discrete time series of either local, global or spatial fields.
A variety of time-dependent external fields are provided, including sinusoidal, sawtooth and rotational fields.
More complex field-protocols can be generated, e.g., for annealing purposes or probing dynamic response.

\subsection{\label{sec:thermal-field}Thermal field}

Thermal fluctuations are modeled as an additional local field, $\vec{h}_\text{th}^{(i)}$, applied to each magnet individually. 
Two orthogonal components of the field are independently drawn from the Normal distribution $\mathcal{N}(0,\sigma_\text{th}^2)$.
The simulated temperature $T$ is closely related to the value of $\sigma_\text{th}$, where a large $\sigma_\text{th}$ corresponds to higher temperatures.

\begin{figure*}[t]
    \subfloat[\label{fig:h-par-perp}] {
        \def\svgwidth{0.25\textwidth}
\begingroup%
  \makeatletter%
  \providecommand\color[2][]{%
    \errmessage{(Inkscape) Color is used for the text in Inkscape, but the package 'color.sty' is not loaded}%
    \renewcommand\color[2][]{}%
  }%
  \providecommand\transparent[1]{%
    \errmessage{(Inkscape) Transparency is used (non-zero) for the text in Inkscape, but the package 'transparent.sty' is not loaded}%
    \renewcommand\transparent[1]{}%
  }%
  \providecommand\rotatebox[2]{#2}%
  \newcommand*\fsize{\dimexpr\f@size pt\relax}%
  \newcommand*\lineheight[1]{\fontsize{\fsize}{#1\fsize}\selectfont}%
  \ifx\svgwidth\undefined%
    \setlength{\unitlength}{255.59999657bp}%
    \ifx\svgscale\undefined%
      \relax%
    \else%
      \setlength{\unitlength}{\unitlength * \real{\svgscale}}%
    \fi%
  \else%
    \setlength{\unitlength}{\svgwidth}%
  \fi%
  \global\let\svgwidth\undefined%
  \global\let\svgscale\undefined%
  \makeatother%
  \begin{picture}(1,0.52221974)%
    \lineheight{1}%
    \setlength\tabcolsep{0pt}%
    \put(0,0){\includegraphics[width=\unitlength,page=1]{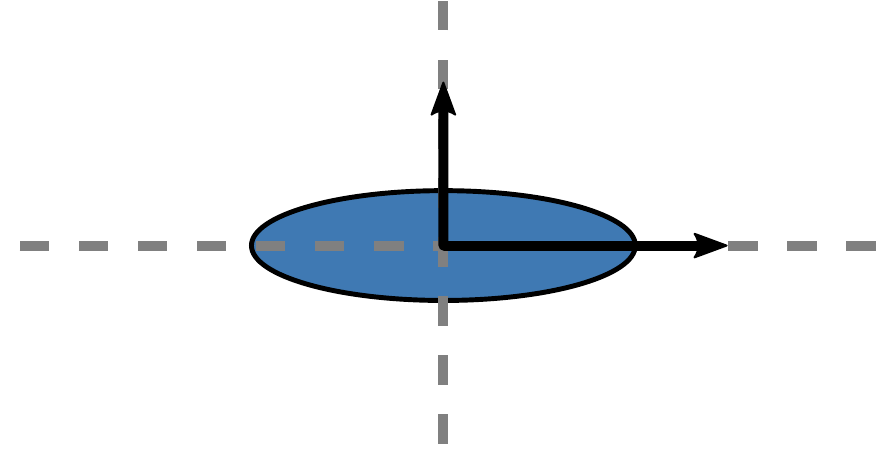}}%
    \put(0.71890087,0.17853772){\color[rgb]{0,0,0}\makebox(0,0)[lt]{\lineheight{0.64999998}\smash{\begin{tabular}[t]{l}$\vec{h}_\parallel$\end{tabular}}}}%
    \put(0,0){\includegraphics[width=\unitlength,page=2]{h_par_perp-ellipsoid.pdf}}%
    \put(0.75676677,0.33035709){\color[rgb]{0,0,0}\makebox(0,0)[lt]{\lineheight{0.64999998}\smash{\begin{tabular}[t]{l}$\vec{h}_i$\end{tabular}}}}%
    \put(0.40407374,0.36647365){\color[rgb]{0,0,0}\makebox(0,0)[lt]{\lineheight{0.64999998}\smash{\begin{tabular}[t]{l}$\vec{h}_\perp$\end{tabular}}}}%
    \put(0.52552927,0.07072981){\color[rgb]{0.4,0.4,0.4}\makebox(0,0)[lt]{\lineheight{0.64999998}\smash{\begin{tabular}[t]{l}hard axis\end{tabular}}}}%
    \put(0.01496577,0.2827307){\color[rgb]{0.4,0.4,0.4}\makebox(0,0)[lt]{\lineheight{0.64999998}\smash{\begin{tabular}[t]{l}easy axis\end{tabular}}}}%
  \end{picture}%
\endgroup%

    }
    \hspace{.2\textwidth}
    \subfloat[\label{fig:h-par-perp-2}]{
        \def\svgwidth{0.25\textwidth}
\begingroup%
  \makeatletter%
  \providecommand\color[2][]{%
    \errmessage{(Inkscape) Color is used for the text in Inkscape, but the package 'color.sty' is not loaded}%
    \renewcommand\color[2][]{}%
  }%
  \providecommand\transparent[1]{%
    \errmessage{(Inkscape) Transparency is used (non-zero) for the text in Inkscape, but the package 'transparent.sty' is not loaded}%
    \renewcommand\transparent[1]{}%
  }%
  \providecommand\rotatebox[2]{#2}%
  \newcommand*\fsize{\dimexpr\f@size pt\relax}%
  \newcommand*\lineheight[1]{\fontsize{\fsize}{#1\fsize}\selectfont}%
  \ifx\svgwidth\undefined%
    \setlength{\unitlength}{255.59999657bp}%
    \ifx\svgscale\undefined%
      \relax%
    \else%
      \setlength{\unitlength}{\unitlength * \real{\svgscale}}%
    \fi%
  \else%
    \setlength{\unitlength}{\svgwidth}%
  \fi%
  \global\let\svgwidth\undefined%
  \global\let\svgscale\undefined%
  \makeatother%
  \begin{picture}(1,0.52221974)%
    \lineheight{1}%
    \setlength\tabcolsep{0pt}%
    \put(0,0){\includegraphics[width=\unitlength,page=1]{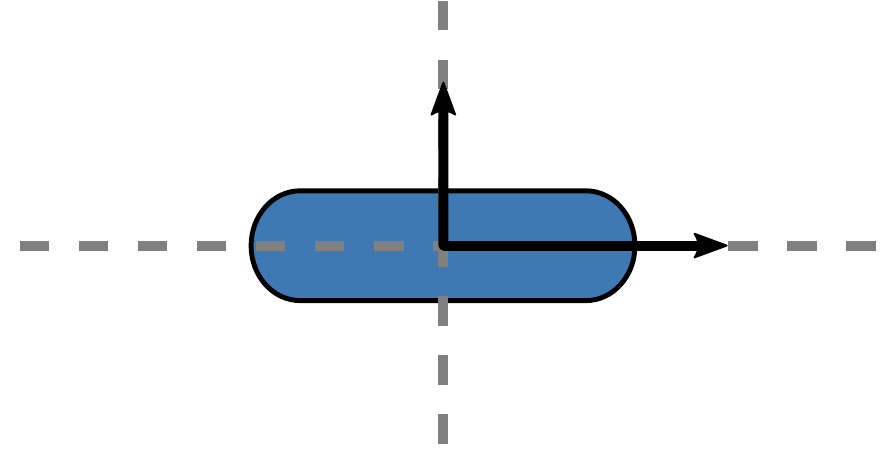}}%
    \put(0.71890087,0.17853768){\color[rgb]{0,0,0}\makebox(0,0)[lt]{\lineheight{0.64999998}\smash{\begin{tabular}[t]{l}$\vec{h}_\parallel$\end{tabular}}}}%
    \put(0,0){\includegraphics[width=\unitlength,page=2]{h_par_perp-updated.pdf}}%
    \put(0.75676677,0.33035705){\color[rgb]{0,0,0}\makebox(0,0)[lt]{\lineheight{0.64999998}\smash{\begin{tabular}[t]{l}$\vec{h}_i$\end{tabular}}}}%
    \put(0.39820519,0.3664736){\color[rgb]{0,0,0}\makebox(0,0)[lt]{\lineheight{0.64999998}\smash{\begin{tabular}[t]{l}$\vec{h}_\perp$\end{tabular}}}}%
    \put(0.52552927,0.07072977){\color[rgb]{0.4,0.4,0.4}\makebox(0,0)[lt]{\lineheight{0.64999998}\smash{\begin{tabular}[t]{l}hard axis\end{tabular}}}}%
    \put(0.00322867,0.28273065){\color[rgb]{0.4,0.4,0.4}\makebox(0,0)[lt]{\lineheight{0.64999998}\smash{\begin{tabular}[t]{l}easy axis\end{tabular}}}}%
  \end{picture}%
\endgroup%

    }
    
    \subfloat[\label{fig:sw-ellipse}] {
        \includegraphics{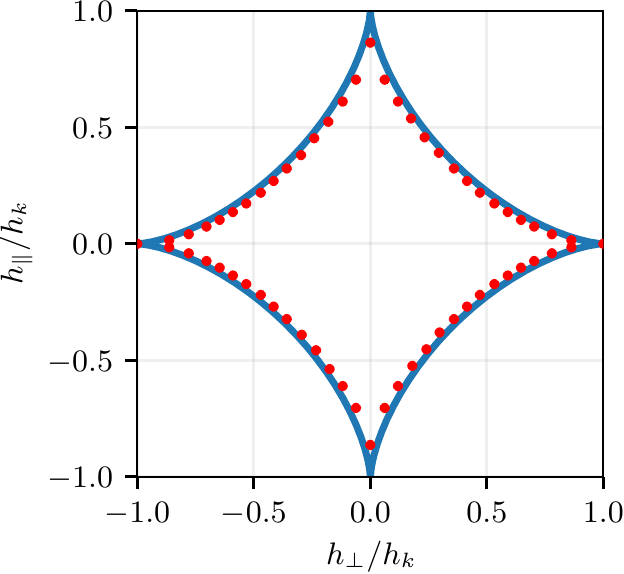}
        \hspace{.06\textwidth}
    } 
    \hspace{.04\textwidth}
    \subfloat[\label{fig:esw-rect}]{
        \includegraphics{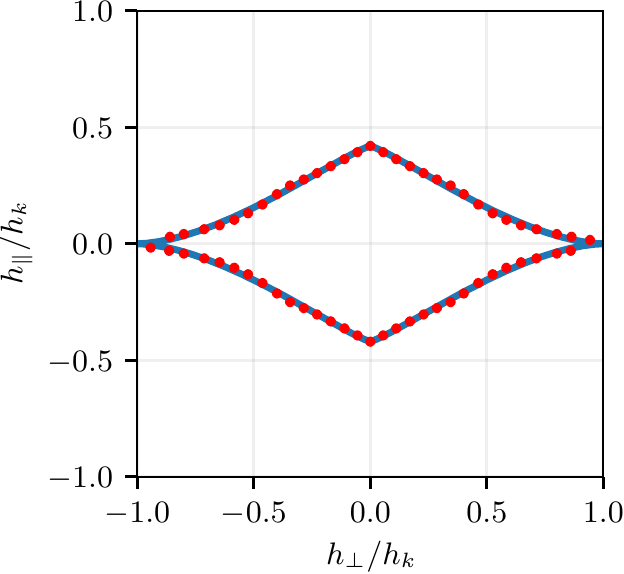}
        \hspace{.06\textwidth}
    }
    
    \caption{\label{fig:astroids}%
    Top: Schematic showing hard and easy axes of (a) an elliptical magnet and (b) a rectangular stadium-shaped magnet, as well as the total field acting on the magnet, $\vec{h}_i$, with its parallel and perpendicular components, $\vec{h}_\parallel$ and $\vec{h}_\perp$, respectively.
    Bottom: Switching astroid for (c) an elliptical magnet and (d) a rectangular stadium-shaped magnet.
    Red dots show the coercive field obtained from micromagnetic simulations.
    The blue line in (c) shows the Stoner-Wohlfarth astroid.
    The blue line in (d) shows the generalized Stoner-Wohlfarth astroid with parameters $b=0.42$, $c=1$, $\beta=1.7$, and $\gamma=3.4$ in \cref{eq:esw}.
    The astroids have been normalized with respect to $h_k$. 
    }
\end{figure*}

When the material and geometric properties of the magnetic islands are known, it is possible to choose a $\sigma_\text{th}$ to match a desired thermal behavior.
In some cases, such as for no thermal field and for thermal protocols with no absolute reference point, it is useful to note that   $\sigma_\text{th}=0$ implies $T=0$, and that $T$ is a monotonically increasing function of $\sigma_\text{th}$.

\subsection{\label{sec:switching}Switching}

Magnetization reversal, or \emph{switching}, may take place when a magnet is subjected to a magnetic field or high temperature.
If the field is sufficiently strong (stronger than some critical field) and directed so that the projection onto $\vec{m}_i$ is in the opposite direction to $\vec{m}_i$, the magnetization $\vec{m}_i$ will switch direction.

The critical field strength is referred to as the coercive field $h_\text{c}$.
For elongated magnets, $h_\text{c}$ depends on the angle between the applied field $\vec{h}_i$ and $\vec{m}_i$.
As illustrated in \cref{fig:h-par-perp}, the \emph{easy axis}, where the magnetization favors alignment, lies along the long axis of the magnet, whereas the \emph{hard axis} is perpendicular to the long axis.
The external field can be decomposed into two components, $\vec{h}_\parallel$ and $\vec{h}_\perp$, corresponding to the field component parallel and perpendicular to the easy axis, respectively.
We denote the coercive field strength along the hard axis as $h_k$. 

A \emph{switching astroid} is a polar plot of $h_\text{c}$ at different angles, with $h_\perp$ on the horizontal axis and $h_\parallel$ on the vertical axis.
It is a description of $\vec{h}_\text{c}$, or $h_\text{c}(\phi)$.
For any applied field $\vec{h}_i$ that is outside the switching astroid, the magnet will switch (given that the projection of $\vec{h}_i$ onto $\vec{m}_i$ is oppositely aligned with respect to $\vec{m}_i$).

\Cref{fig:sw-ellipse} shows the normalized switching astroid for an elliptical magnet (\cref{fig:h-par-perp}) as obtained from micromagnetic simulations (red dots).
Notice how $h_\text{c}$ is the smallest at a $45\degree$ angle, indicating that a field directed at \SI{45}{\degree} to a magnet's principal axes will require the least field strength in order to switch its magnetization.

The Stoner-Wohlfarth (SW) model captures the angle dependent switching characteristic of single-domain elliptical magnets\cite{stoner-wohlfarth-Tannous_2008}. 
The characteristic SW astroid is shown in \cref{fig:sw-ellipse} (blue line) and is described by the equation
\begin{equation} \label{eq:sw}
    \left(\frac{h_\parallel}{h_k}\right)^{2/3} + 
    \left(\frac{h_\perp}{h_k}\right)^{2/3} = 1.
\end{equation}
In the SW model, switching may occur when the left hand side of \cref{eq:sw} is greater than one.

The astroid obtained from micromagnetic simulations and the SW astroid (\cref{fig:sw-ellipse}) are nearly identical, indicating that the SW model is a simple and valid description of switching in elliptical nanomagnets. 

However, the SW model is only accurate for elliptical magnets, other magnet shapes typically have quite different switching characteristics.
\Cref{fig:esw-rect} shows the switching astroid for rectangular stadium-shaped magnets (red dots), which is the shape commonly used in most fabricated ASIs (\cref{fig:h-par-perp-2}).
Notice how the astroid is asymmetric: rectangular magnets switch more easily along the easy axis than the hard axis.

To capture the asymmetric switching characteristics of non-elliptical magnets, we have generalized the SW switching model to allow an asymmetry between easy and hard axes. 
Additionally, the model has been extended to support tuning of the curvature of the extrema.
In the generalized model, the switching threshold is given by
\begin{equation} \label{eq:esw}
    \left(\frac{h_{\parallel}}{b h_k}\right)^{2/\gamma} + 
    \left(\frac{h_{\perp}}{c h_k}\right)^{2/\beta} = 1,
\end{equation}

where $b$, $c$, $\beta$ and $\gamma$ are parameters which adjust the shape of the astroid: $b$ and $c$ define the height and width, respectively, while $\beta$ and $\gamma$ adjust the curvature of the astroid at the easy and hard axis, respectively. 
Introducing these new parameters allows for tuning of the switching astroid to fit with the shape of nanomagnets used in ASIs.  
With $b=c=1$ and $\beta=\gamma=3$, \cref{eq:esw} reduces to \cref{eq:sw}, i.e., the classical Stoner-Wohlfarth astroid is obtained (valid for elliptical magnets).

By tuning the parameters of the generalized SW model, we can obtain the asymmetric switching astroid shown in \cref{fig:esw-rect} (blue line).
The astroid is in good agreement with results obtained from micromagnetic simulations (red dots). 

In flatspin, the generalized SW model is used as the switching criteria, i.e., a spin may flip if the left hand side of \cref{eq:esw} is greater than one.
Additionaly, the projection of $\vec{h}_\text{i}$ onto $\vec{m}_i$ must be in the opposite direction of $\vec{m}_i$:

\begin{equation} \label{eq:sw2}
    \vec{h}_i \cdot \vec{m}_i < 0.
\end{equation}

\subsection{\label{sec:disorder}Imperfections and disorder}

Due to manufacturing imperfections there will always be a degree of variation in the shape and edge roughness of nanomagnets. 
This variation can be thought of as a disorder in the magnets' inherent properties. 
Rough edges and a slightly distorted geometry can affect the magnets' switching mechanisms, with defects pinning magnetization and altering the coercive field for each magnet.

In flatspin we model this variation as disorder in the coercive fields.
The coercive field is defined individually for each magnet, and a distribution of values can be used to introduce variation.
A user-defined parameter, $k_\text{disorder}$, defines the distribution of coercive fields, i.e., $h_{k}^{(i)}$ is sampled from a normal distribution $\mathcal{N}(h_{k}, \sigma)$, where $\sigma = k_\text{disorder} \cdot h_k$ (while ensuring $h_{k}^{(i)}$ is always positive).

\subsection{\label{sec:dynamics}Dynamics}

flatspin employs deterministic single spin flip dynamics.
At each simulation step, we calculate the total magnetic field, $\vec{h}_i$, acting on each spin.
Next, we determine which spins \textit{may} flip according to the switching criteria \cref{eq:esw,eq:sw2}.
Finally we flip the spin where $\vec{h}_i$ is the furthest outside its switching astroid, i.e., where the left hand side of \cref{eq:esw} is greatest.
Ties are broken in a deterministic, arbitrary manner, although with non-zero disorder such occurrences are rare.
The above process is repeated until there are no more flippable spins.

This relaxation process is performed with constant external and thermal fields.
To advance the simulation, the fields are updated and relaxation is performed again.
Hence a simulation run consists of a sequence of field updates and relaxation processes.

The dynamical process makes three main assumptions:
\begin{enumerate}
    \item The external field is quasi-static compared to the time scale of magnet switching.
    \item Magnet switching is sequential.
    \item The magnet experiencing the highest effective field compared to its switching threshold is the first to switch. 
\end{enumerate}

\textit{Assumption 1} means the model holds for low frequency external fields, i.e., where switching settles under a static field.
The switching mechanics of nanomagnets are typically in the sub nanosecond range \cite{Kikuchi1956-nanoswitching, Gillette1958-nanoswitching}, and experimental setups often employ external magnetic fields which can be considered static at this time scale.
At high applied field frequencies, more complex physical phenomena such as spin waves will have a non-negligible effect on switching dynamics, which are not modeled in flatspin.

\textit{Assumption 2} holds if the coercive fields $h_\text{c}^{(i)}$, and total field $\vec{h}_i$, of the magnets are sufficiently non-uniform, so that there will always be a single magnet which will flip first.
It is assumed to be unlikely that two magnets will have the same $h_\text{c}^{(i)}$ \emph{and} $\vec{h}_i$ simultaneously.
However, in those rare cases where two magnets are equally close to switching, overlapping switching events may occur in a physical system.

\textit{Assumption 3} relies on the fact that all changes in the magnetic fields are effectively continuous, and the change is unidirectional within a simulated time step, i.e., a \emph{quasi-static field}.
Since evaluation happens in discrete time, there will be cases where several magnets are above their corresponding switching thresholds simultaneously.
In those cases, the magnet furthest above its switching threshold will have been the first to have crossed the threshold under a quasi-static field.
Furthermore, if the angle of the external field is constant, the switching order is invariant to the time resolution of the external field.

\begin{figure*}
    \subfloat[\label{fig:geom-square-closed}]{
        \centering
        \includegraphics{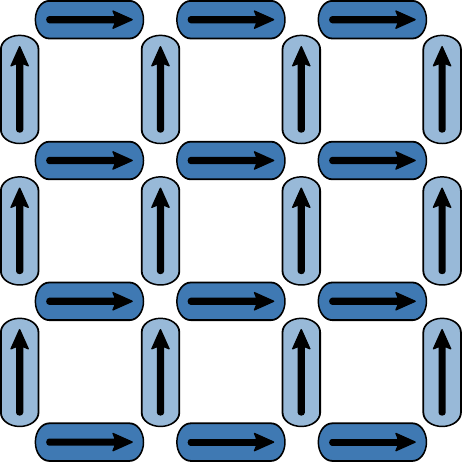}
    }
    \hspace{.3in}%
    \subfloat[\label{fig:geom-square-open}]{
        \centering
        \includegraphics{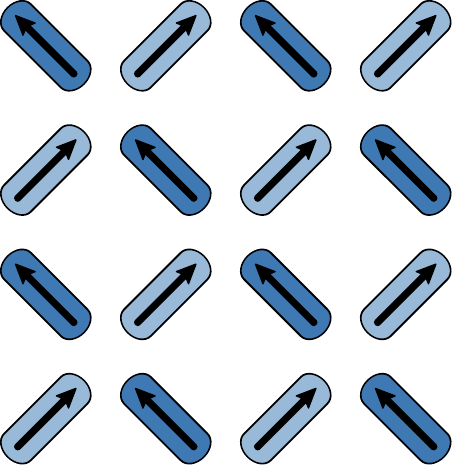}
    }
    \hspace{.3in}%
    \subfloat[\label{fig:geom-kagome}]{
        \centering
        \includegraphics{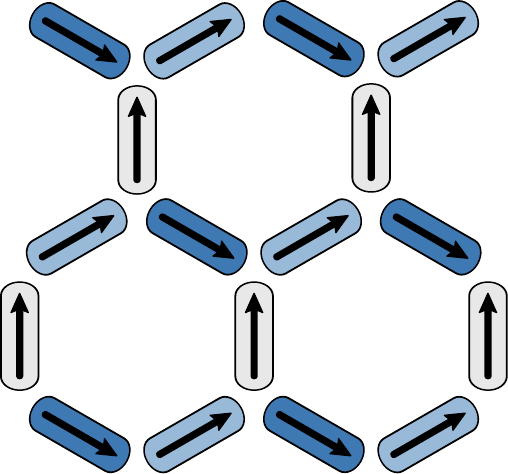}
    }
    
    \subfloat[\label{fig:geom-pinwheel-diamond}]{
        \centering
        \includegraphics{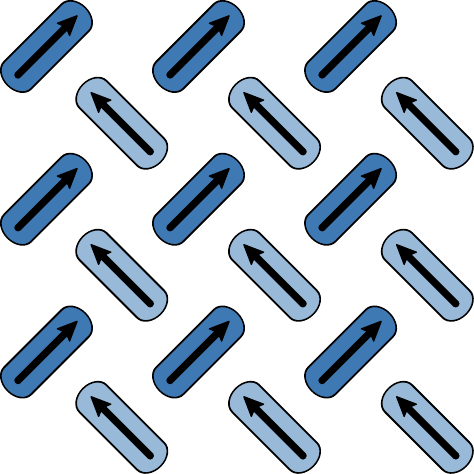}
    }
    \hspace{.3in}%
    \subfloat[\label{fig:geom-pinwheel-lucky-knot}]{
        \centering
        \includegraphics{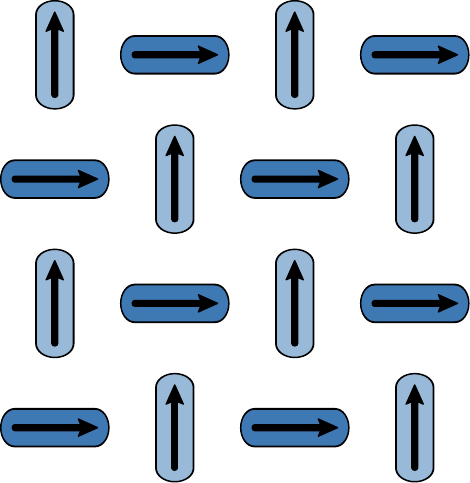}
    }
    \hspace{.3in}%
    \subfloat[\label{fig:geom-spin-ice}] {
        \centering
        \includegraphics{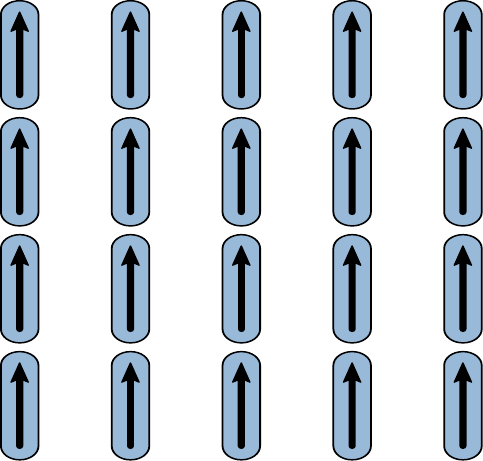}
    }
    \caption{\label{fig:geometries}%
    flatspin includes the most common ASI geometries: (a) Square (closed edges), (b) Square (open edges), (c) Kagome, (d) Pinwheel ``diamond'', (e) Pinwheel ``lucky-knot'', and (f) Ising.
    }
\end{figure*}

\subsection{\label{sec:geometries}Geometries}

The particular spatial arrangement of the magnets is referred to as the \emph{geometry}. %
A wide range of ASI geometries have been proposed in the literature.
\Cref{fig:geometries} depicts the geometries included in flatspin, which are the most commonly used ASI geometries:  square\citep{Wang2006}, kagome\citep{Tanaka2006,Qi2008}, pinwheel\citep{Macedo2018,Li2019} and ising\citep{random-ising-farhan}.

Geometries are often decomposed into two or more ``sublattices'', where the magnets within one sublattice are all aligned (have the same rotation).
In \cref{fig:geometries}, the sublattice a magnet belongs to is indicated by its color.
As can be seen, both square and pinwheel ASIs have two perpendicular sublattices, whereas kagome has three sublattices.

flatspin can be used to model any two-dimensional ASI comprised of identical elements.
New geometries can easily be added by extending the model with a new set of positions $\vec{r}_i$ and rotations $\theta_i$.

\section{\label{sec:simulation_framework}Simulation framework}

In addition to a magnetic model, flatspin provides a flexible framework for running simulations, storing results, and performing analysis.

\begin{figure}
    \centering
    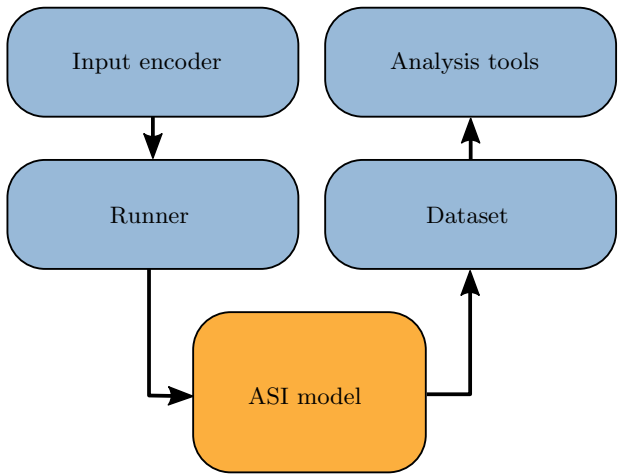
    \caption{\label{fig:sys-arch-overview}%
    Overview of the flatspin architecture, with arrows indicating data flow.
    }
\end{figure}

\Cref{fig:sys-arch-overview} illustrates the overall architecture of flatspin.
The \emph{ASI model} has been described in detail in \cref{sec:modeling}.
Conceptually, the ASI model describes the physical system under study.
The rest of the components are tools used by the experimenter to interact with the ASI and observe the results.
In this section we briefly describe each of these components.

The \emph{input encoder} translates a set of input values to a series of external fields.
Encoders provide a flexible way to define field protocols, and have been designed with neuromorphic computing in mind.
A range of encoders are included, e.g., sinusodial, sawtooth and rotational fields.

The responsibility of the \emph{runner} component is to perturb the ASI model according to the field protocol, and save the results.
The model, which is fully parametric, receives parameters from the runner, enabling automated parameter sweeps.
In addition, there is support for distributed running of simulations on a compute cluster.

Results are stored in a well-defined \emph{dataset} format which makes the analysis of a large numbers of simulations straightforward.
A suite of \emph{analysis tools} are included, e.g., for plotting results, visualizing ensemble dynamics and analysis of vertex populations.

flatspin is written in Python and utilizes OpenCL to accelerate calculations on the GPU.
OpenCL is supported by most GPU vendors, hence flatspin can run accelerated on a wide variety of platforms.
The simulator may also run entirely on CPUs in case GPUs are not available, albeit at a reduced speed.

flatspin is open-source software and released under a GNU GPL license.
For more information see the website\citep{FlatspinWebsite} and User Manual \citep{FlatspinUserManual}.

\section{Validation of flatspin}

To evaluate the ASI model, flatspin simulations were compared to established experimental results from the literature, as well as micromagnetic simulations.
In particular we investigate phenomena such as Dirac strings in kagome ASI, the size of crystallite domains in square ASI, and superferromagnetism in pinwheel ASI.
Finally, we compare the switching order from flatspin simulations with that of micromagnetic simulations, and investigate the effect of varying lattice spacings.

\subsection{Dirac strings in kagome ASI}

To assess the ability of flatspin to reproduce fine-scale patterns, we consider the emergence of Dirac strings in a kagome ASI (\cref{fig:geom-kagome}). Applying a reversal field to a polarized kagome ASI results in the formation of monopole-antimonopole pairs \citep{mengotti2011}. These pairs are joined by a ``string'' of nanomagnets which have flipped due to the reversal field. As the strength of the reversal field increases, the strings elongate until they fill the array. 

We closely follow the methodology set out in an experimental study of Dirac stings in kagome ASI \cite{mengotti2011}, in which a room temperature kagome ASI undergoes magnetization reversal. We start with an array of $2638$ magnets ($29\times29$ hexagons) polarized to the left and apply a reversal field $\vec{H}$ to the right with a slight, downward offset of $3.6\degree$. This offset breaks the symmetry, such that one of the sublattices is now least aligned with the field, resulting in an increased coercive field on this ``unfavored'' sublattice. 

Micromagnetic simulations of magnets of size \SI{470x160x20}{\nano\meter} yield the following estimation of flatspin parameters: $\alpha=0.00103, h_k=0.216, \beta=2.5, \gamma=3,  b=0.212, c=1$. The temperature and disorder parameters $\sigma_\text{th}=0.002$ and $k_\text{disorder}=0.05$ were determined empirically through qualitative comparison with the experimental results \cite{mengotti2011}.

\begin{figure*}[ht]
    \centering
    \includegraphics{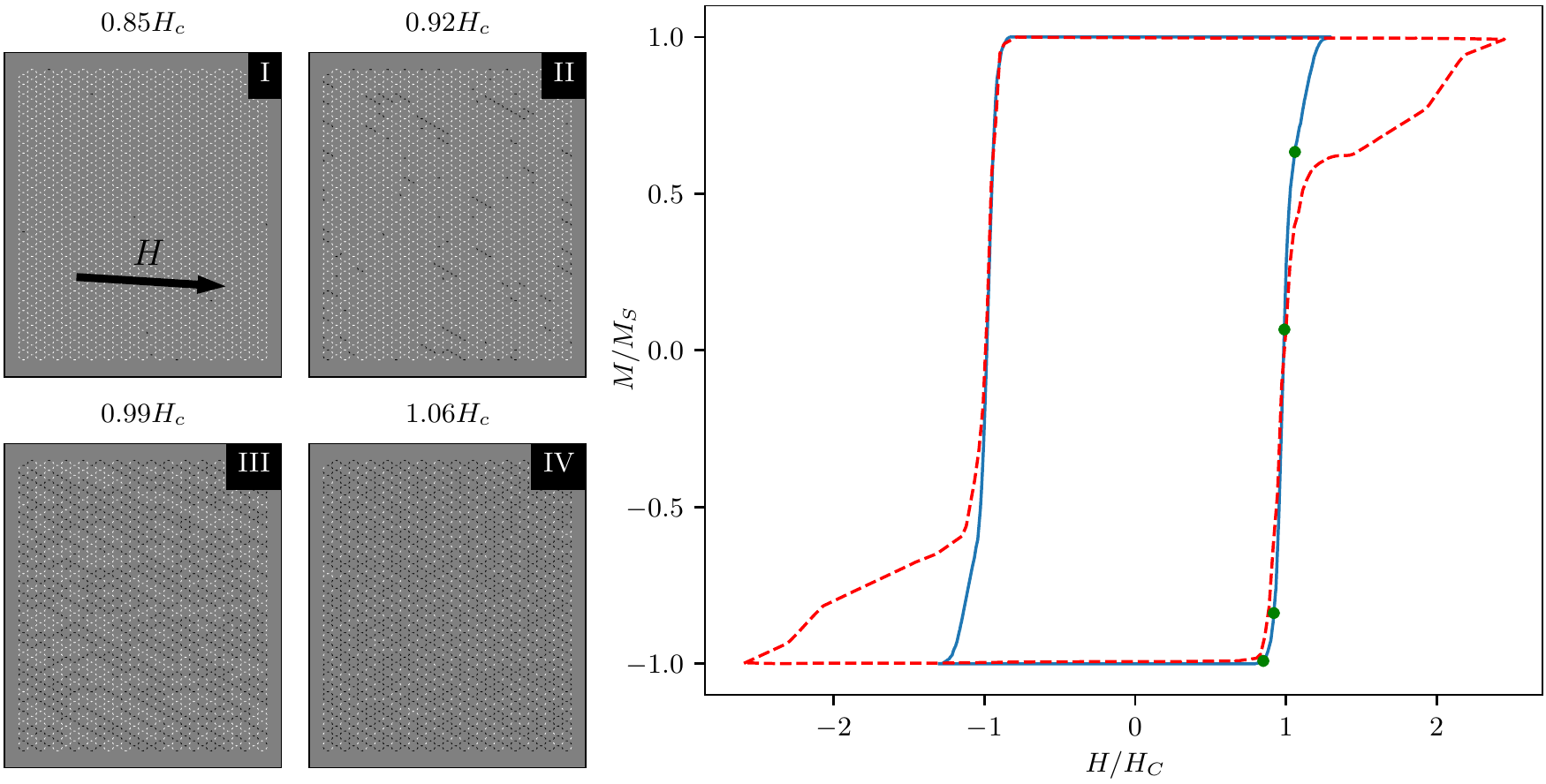}
    \caption{\label{fig:dirac-strings}%
    Left: snapshots of the evolution of a kagome ASI at selected field values.
    Right: Comparison of the hysteresis curve of the simulated ensemble (blue line) against a sketch of the hysteresis curve from the experimental results \cite{mengotti2011} (red dashed line). Green dots indicate the points at which the snapshots are sampled from.
    }
\end{figure*}

The time evolution snapshots of \cref{fig:dirac-strings} demonstrate a strong, qualitative similarity to the results of \textcite{mengotti2011}. We see Dirac strings developing with a preference to lie along the two sublattices most aligned with the field angle.
Furthermore, in the final image, we see the vast majority of unflipped magnets (excluding the edges) are on the unfavored sublattice, in accordance with both experimental and simulated results from the literature.

Also in \cref{fig:dirac-strings}, we see the hysteresis of the simulated ensemble (solid line) is similar to that of \textcite{mengotti2011} (dashed line) in some sections, but differs near the extrema. The hysteresis can be understood in two phases. The first phase, at roughly $M/M_\text{S} \in [-0.6,0.6]$, is dominated by the lengthening of the Dirac strings, with almost no activity occurring on the unfavored sublattice. At $M/M_\text{S} < -0.6$ and $M/M_\text{S} > 0.6$, the ensemble enters a second phase in which the Dirac strings have fully covered the array, and change in magnetization is dominated by switching on the unfavored sublattice. Clearly we see good agreement, within phase one, between our simulated hysteresis and the experimental results. Furthermore, there is a clear phase transition (characterized by a sharp decrease in gradient) in our hysteresis very close to the transition in the experimental hysteresis. Notably however, although the phase transitions occur at a similar time, the change in gradient is less pronounced in our simulated hysteresis. This disparity indicates that, in the second phase, the magnets on the unfavored sublattice flip more easily in our simulation than in the experimental data.

The accuracy of the point dipole approximation is known to suffer when considering kagome ASI. Specifically, it has been shown to underestimate the coupling coefficient of the nearest neighbors by approximately a factor of $5$ \cite{rougemaille2011}. Despite this, we observe flatspin accurately reproduces snapshots of the time evolution behavior, while also capturing salient features of the ensemble hysteresis curve.

\subsection{Domain size in square ASI}
In order to demonstrate simulation of large-scale behavior, we have reproduced the emergence of large domains of magnetic order in square ASI, similar to experimental results of \textcite{Zhang2013}.
One of the main advantages of flatspin over typical alternatives is the scalability and high throughput of large systems with many magnets. 
Some emergent ASI phenomena require large systems in order to be fully quantified and studied with high fidelity, such as the domain size of magnetic charge crystallites.
For ASIs with strongly coupled magnets, typical domain sizes can become too large for direct experimental observation.
Thus, an accurate estimate of the domain size for ASIs with a small lattice constant is, in part, limited by the number of directly observable magnets. 

For a given range of lattice constants covering both strongly coupled ASIs and weakly coupled ASIs, a corresponding range of large to small magnetic order coherence lengths is expected.  
In this study, we consider square ASI (closed edges, \cref{fig:geom-square-closed}) with different lattice constants, $a$, ranging from \SIrange{320}{880}{\nano\meter}.

$50 \times 50$ square ASIs were annealed in flatspin using a thermal protocol of exponentially decreasing $\sigma_\text{th}$.
A switching astroid for \SI{220x80x25}{\nano\meter} was obtained through micromagnetic simulations, described by generalized astroid parameters $b = 0.4$, $c = 1.0$, $\beta = 3.0$, and $\gamma = 3.0$.
Additionally, $h_k=0.186$, $k_\text{disorder}=0.05$, and a neighbor distance of 10 magnets was used.
The thermal protocol was chosen such that the total dipole interaction energy was not significantly reduced by increasing simulation steps.

\begin{figure*}[t]
    \subfloat[\label{fig:crystallite-correlation-sim-b}]{
    \centering
        \def\svgwidth{0.45\linewidth}   
\begingroup%
  \makeatletter%
  \providecommand\color[2][]{%
    \errmessage{(Inkscape) Color is used for the text in Inkscape, but the package 'color.sty' is not loaded}%
    \renewcommand\color[2][]{}%
  }%
  \providecommand\transparent[1]{%
    \errmessage{(Inkscape) Transparency is used (non-zero) for the text in Inkscape, but the package 'transparent.sty' is not loaded}%
    \renewcommand\transparent[1]{}%
  }%
  \providecommand\rotatebox[2]{#2}%
  \newcommand*\fsize{\dimexpr\f@size pt\relax}%
  \newcommand*\lineheight[1]{\fontsize{\fsize}{#1\fsize}\selectfont}%
  \ifx\svgwidth\undefined%
    \setlength{\unitlength}{340.64001465bp}%
    \ifx\svgscale\undefined%
      \relax%
    \else%
      \setlength{\unitlength}{\unitlength * \real{\svgscale}}%
    \fi%
  \else%
    \setlength{\unitlength}{\svgwidth}%
  \fi%
  \global\let\svgwidth\undefined%
  \global\let\svgscale\undefined%
  \makeatother%
  \begin{picture}(1,1.06750652)%
    \lineheight{1}%
    \setlength\tabcolsep{0pt}%
    \put(0,0){\includegraphics[width=\unitlength,page=1]{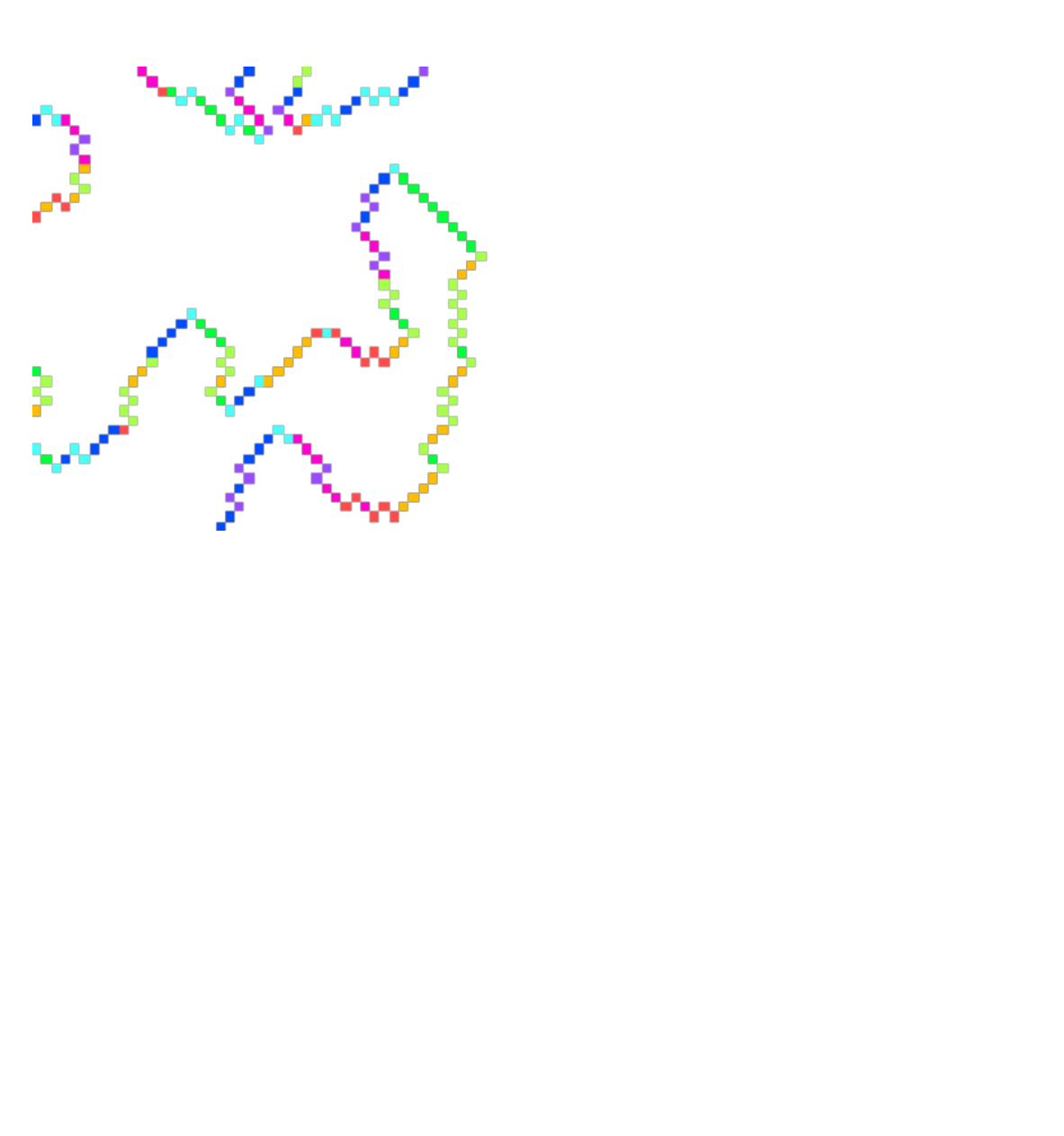}}%
    \put(0.24821681,1.01941805){\color[rgb]{0,0,0}\makebox(0,0)[t]{\lineheight{1.25}\smash{\begin{tabular}[t]{c}$a = $\SI{360}{\nano\meter}\end{tabular}}}}%
    \put(0,0){\includegraphics[width=\unitlength,page=2]{50x50_domains-color.pdf}}%
    \put(0.75188636,0.48236134){\color[rgb]{0,0,0}\makebox(0,0)[t]{\lineheight{1.25}\smash{\begin{tabular}[t]{c}$a = $\SI{480}{\nano\meter}\end{tabular}}}}%
    \put(0,0){\includegraphics[width=\unitlength,page=3]{50x50_domains-color.pdf}}%
    \put(0.24743172,0.48236134){\color[rgb]{0,0,0}\makebox(0,0)[t]{\lineheight{1.25}\smash{\begin{tabular}[t]{c}$a = $\SI{440}{\nano\meter}\end{tabular}}}}%
    \put(0,0){\includegraphics[width=\unitlength,page=4]{50x50_domains-color.pdf}}%
    \put(0.75188637,1.01941805){\color[rgb]{0,0,0}\makebox(0,0)[t]{\lineheight{1.25}\smash{\begin{tabular}[t]{c}$a = $\SI{400}{\nano\meter}\end{tabular}}}}%
    \put(0,0){\includegraphics[width=\unitlength,page=5]{50x50_domains-color.pdf}}%
  \end{picture}%
\endgroup%

    }
    \hspace{.2in}%
    \subfloat[\label{fig:crystallite-correlation-sim-a}]{
        \centering
        \includegraphics[width=0.45\textwidth]{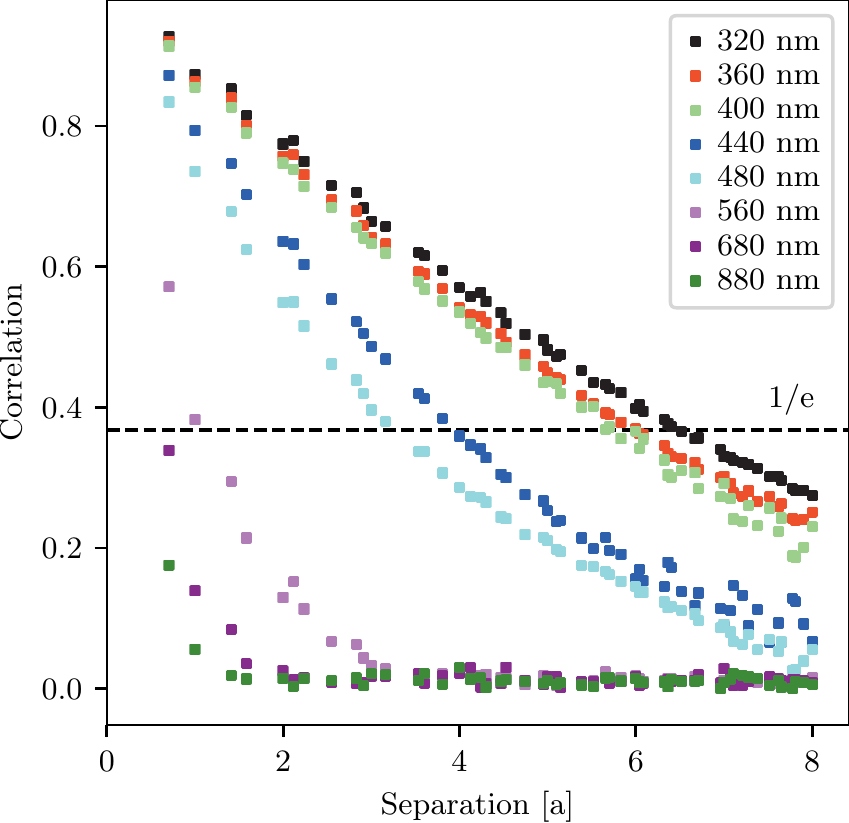}
        }
    \caption{\label{fig:crystallite-correlation}
    (a) Maps showing the net magnetization of the ASI vertices, for an annealed $50\times50$ square ASI with the given lattice constant $a$.
    The white regions have zero net magnetization and thus correspond to a coherent domain of Type I vertices. Colored regions have a non-zero net magnetization, direction indicated by the color wheel, and correspond to Type II or III vertices.
    (b) The absolute value of spin-spin correlation at a given separation for square ASIs of different lattice constants, $a$. 
    Also indicated is a $1/e$ threshold of correlation (dashed line).  
    }
\end{figure*}

In the annealed state, the spin-spin correlation as a function of their lateral separation was calculated across the ensembles.
Analysis of the average correlation of annealed states provides insight about the typical coherence length of magnetic order, i.e., magnetic charge crystallite size, or domain size.
Here, the correlation of two spins is defined as +1 (-1) if their dipole interaction is minimized (maximized). 
Averaging correlation across distinct types of spin pairs, in the annealed ASI, gives a measure of how coherent the ASI is at that particular neighbor separation. 
How quickly the average correlation decreases as a function of separation can be used to estimate the characteristic domain size. 
In particular, it can be argued that the separation where the correlation falls below $1/e$ is the characteristic domain radius~\citep{Newman1999, Zhang2013}. 

Typical domain structures and correlation results can be seen in \cref{fig:crystallite-correlation}. 
The domains shown in \cref{fig:crystallite-correlation-sim-b} and the correlation curves in \cref{fig:crystallite-correlation-sim-a} are, for the most part, in good agreement with experimental results\citep{Zhang2013}.
A qualitative comparison of the domain sizes and structures in \cref{fig:crystallite-correlation} shows that the domains tend to be larger, with smoother domain boundaries, for smaller $a$.
The analysis of coherence as a function of separation also shows identical trends and similar values, where an increase in $a$ leads to low correlation, even between nearest neighbors. 

For cases where $a<$~\SI{400}{\nano\meter}, domains do not significantly increase in size when the lattice constant is further reduced.
This discrepancy with the experimental results is not completely unexpected: the point-dipole approximation is known to underestimate nearest-neighbor interaction for magnets placed close together~\cite{rougemaille2011}. 
In addition, a stronger interaction between spins would cause each spin flip to contribute a greater change in the total dipole energy. 
This makes a gradual descent towards the ground state by random spin flips (the thermal fluctuations as modeled by flatspin) harder to achieve.
These issues may be addressed by increasing the coupling parameter $\alpha$ for nearest neighbor spins, and by a longer and slower annealing protocol. 
A longer and slower annealing protocol will inevitably come at the cost of longer computation times.
In a future version of flatspin, it might be beneficial to include other thermal effects such as a declining saturation magnetization of the constituent material. 

These results show that flatspin provides sufficient flexibility, fidelity and performance required to reproduce experimentally observed large-scale emergent behavior in ASIs.

\subsection{\label{sec:pinwheel}Superferromagnetism in pinwheel ASI}

In this section, we use flatspin to reproduce the dynamic behavior of pinwheel ASI, which had yet to be demonstrated with a dipole model\cite{Li2019}.
We find that our switching criteria plays a key role in replicating magnetization details during the field-driven array reversal.

Pinwheel ASI is obtained by rotating each island in square ASI some angle about its center.
A rotation of 45 degrees results in a transition from antiferromagnetic to ferromagnetic order\citep{Macedo2018}.
The dynamics of pinwheel ASI in many ways resemble continuous ferromagnetic thin films, with mesoscopic domain growth originating from nucleation sites, followed by coherent domain propagation and complete magnetization reversal \citep{Li2019}.

Here we demonstrate that flatspin is able to replicate the experimental reversal processes presented in \textcite{Li2019}, where pinwheel ``diamond'' ASI (\cref{fig:geom-pinwheel-diamond}) is subject to an external field at different angles.
A key result is that the angle $\theta$ of the external field controls the nature of the reversal process.
When $\theta$ is small (equally aligned to both sublattices), reversal happens in a single avalanche, whereas when $\theta$ is large (more aligned to one sublattice), reversal happens in a two-step process where one sublattice switches completely before the other.
Previous attempts at capturing this behavior in a dipole model have proven unfruitful\cite{Li2019}.

To replicate this process in flatspin, an asymmetric switching astroid is required, i.e., the threshold along the parallel component is reduced by setting $b < 1$ in \cref{eq:esw}.
From micromagnetic simulations of a single \SI[product-units = single]{470 x 170 x 10}{\nano\meter} magnet we obtain the following 
characteristic switching parameters: $b = 0.28$, $c = 1.0$, $\beta = 4.8$ and $\gamma = 3.0$.
Other simulation parameters include $\alpha \approx 0.00033$, $h_k=0.098$, $k_\text{disorder}=0.05$ and a neighbor distance of 10.

\begin{figure*}[p]
    \def\pwfigwidth{0.38\textwidth}
    \subfloat[\label{fig:li2019-theta0}]{
        \includegraphics[width=\pwfigwidth]{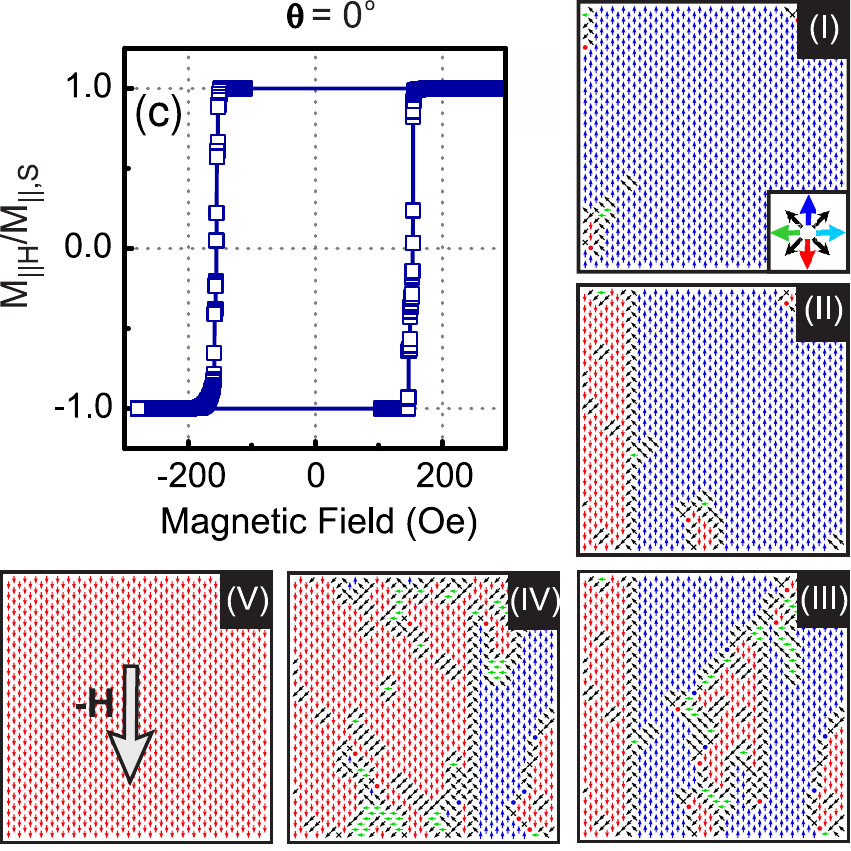}
    }
    \hspace{.4in}%
    \subfloat[\label{fig:pw-theta0}]{
        \includegraphics[width=\pwfigwidth]{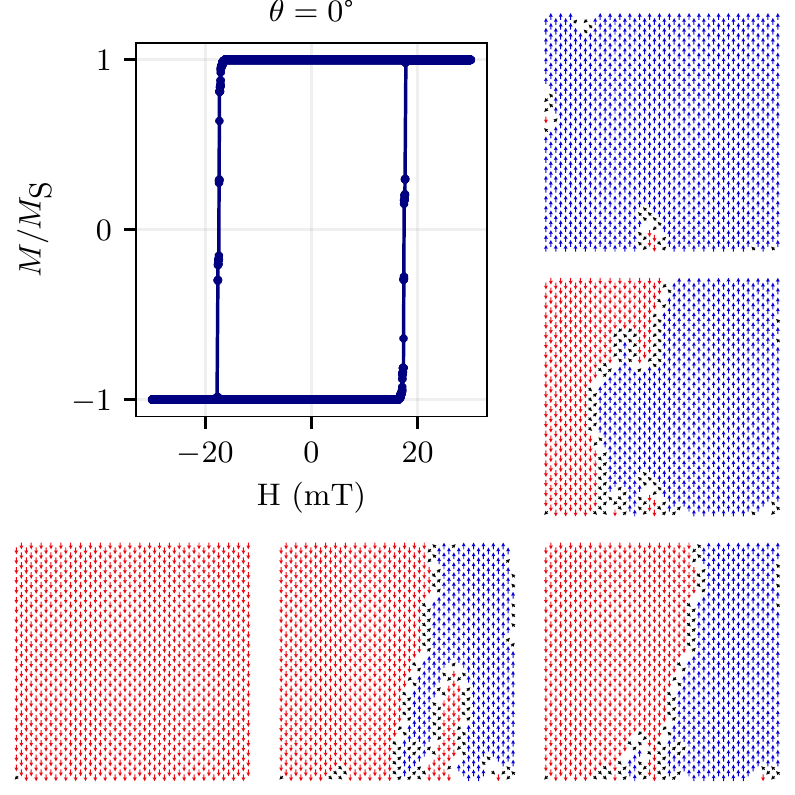}
    }
    
    \subfloat[\label{fig:li2019-theta6}]{
        \includegraphics[width=\pwfigwidth]{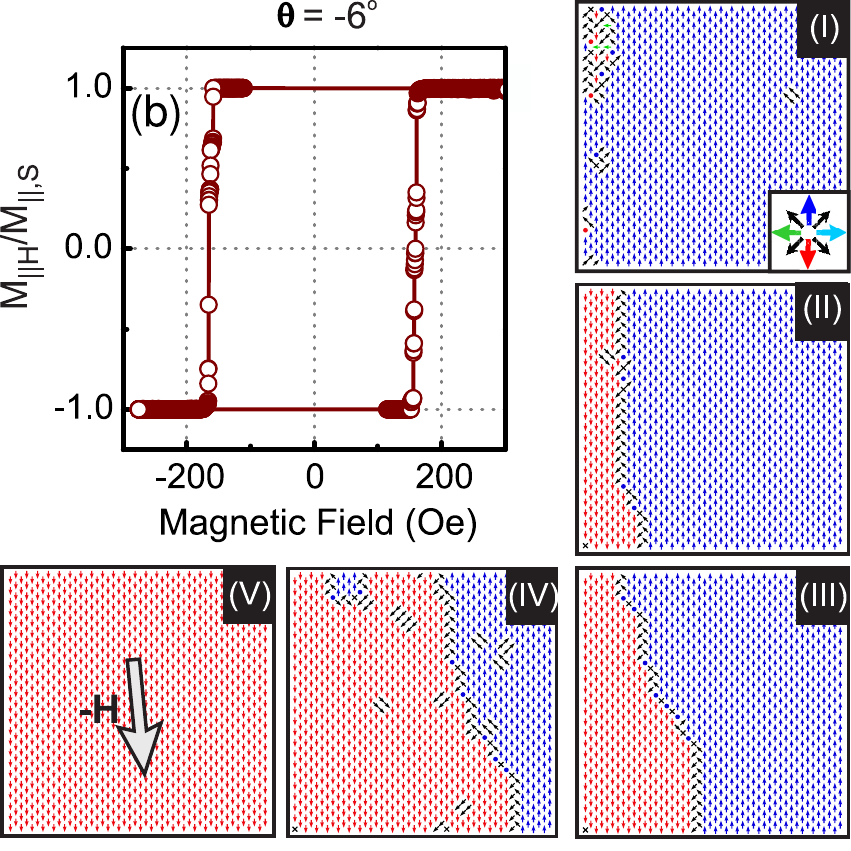}
    }
    \hspace{.4in}%
    \subfloat[\label{fig:pw-theta6}]{
        \includegraphics[width=\pwfigwidth]{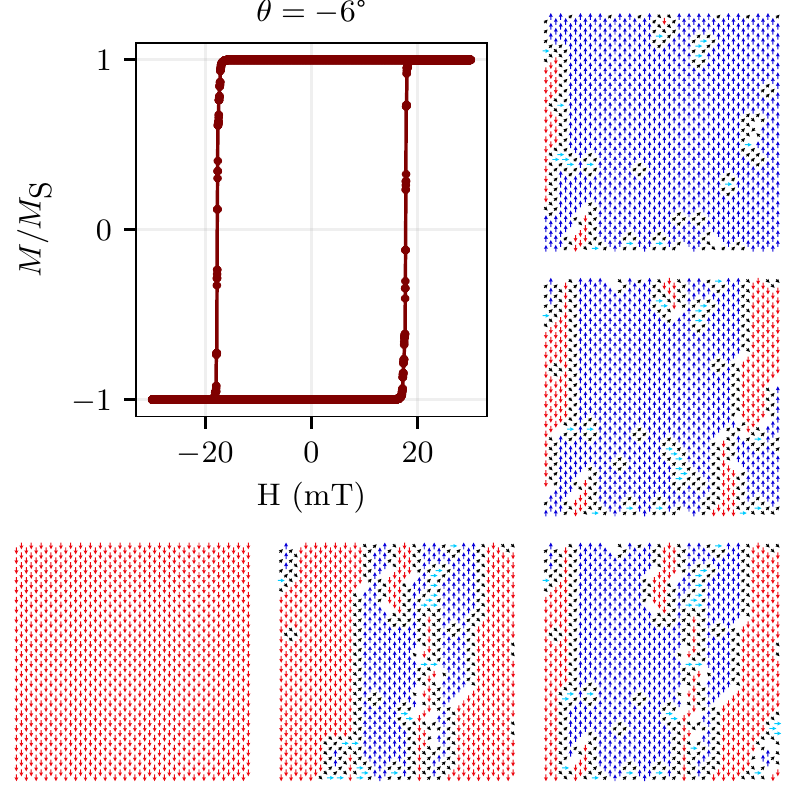}
    }
    
    \subfloat[\label{fig:li2019-theta30}]{
        \includegraphics[width=\pwfigwidth]{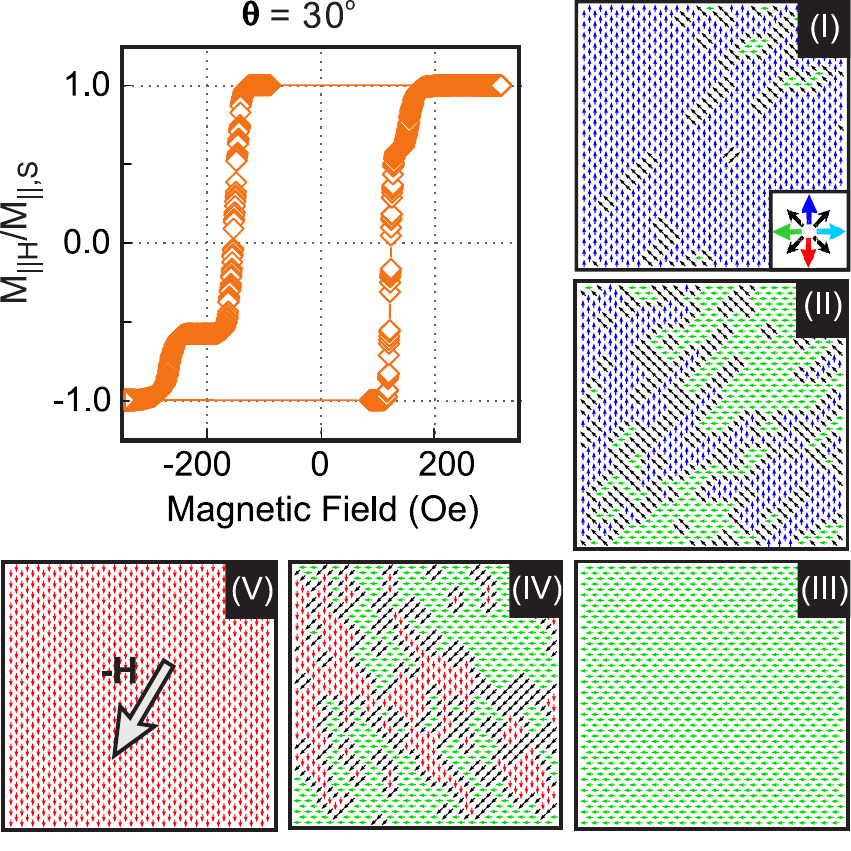}
    }
    \hspace{.4in}%
    \subfloat[\label{fig:pw-theta30}]{
        \includegraphics[width=\pwfigwidth]{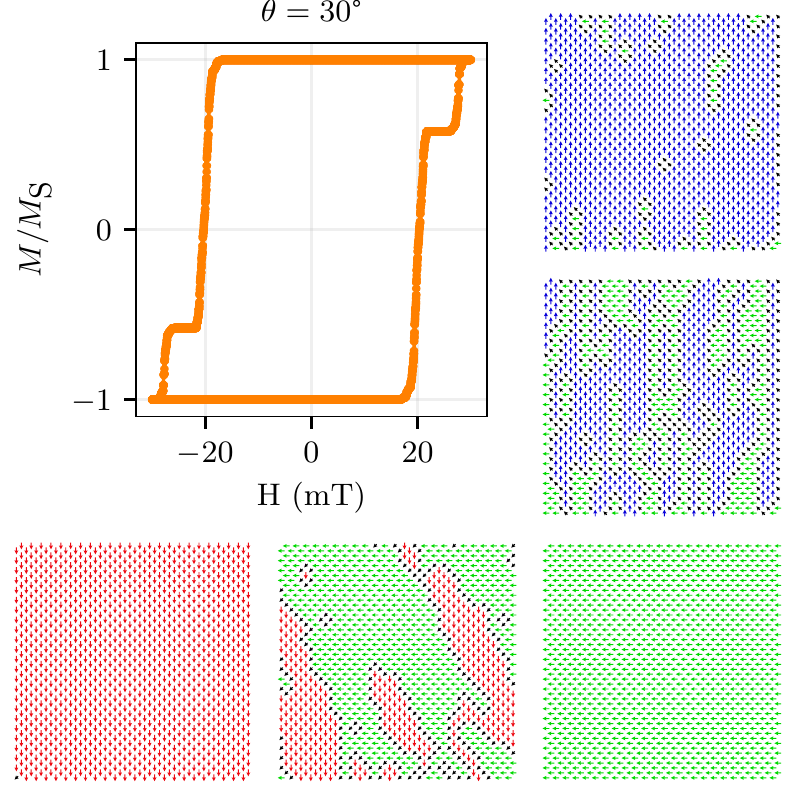}
    }
    \caption{\label{fig:pinwheel-theta0}%
    Hysteresis loop and snapshots of the pinwheel units for various angles $\theta$ of the applied field.
    Figures (a), (c) and (e) show experimental results, adapted from \textcite{Li2019}, Copyright \textcopyright 2018 American Chemical Society, CC-BY-4.0 \url{https://creativecommons.org/licenses/by/4.0/}.
    Figures (b), (d) and (f) show results from flatspin simulation (white indicates zero net magnetization).
    }
\end{figure*}
\clearpage

\begin{figure*}[tp!]
    \subfloat[\label{fig:fs_mm_spearmana}] {
        \includegraphics[height=2.3in,width=0.4\textwidth]{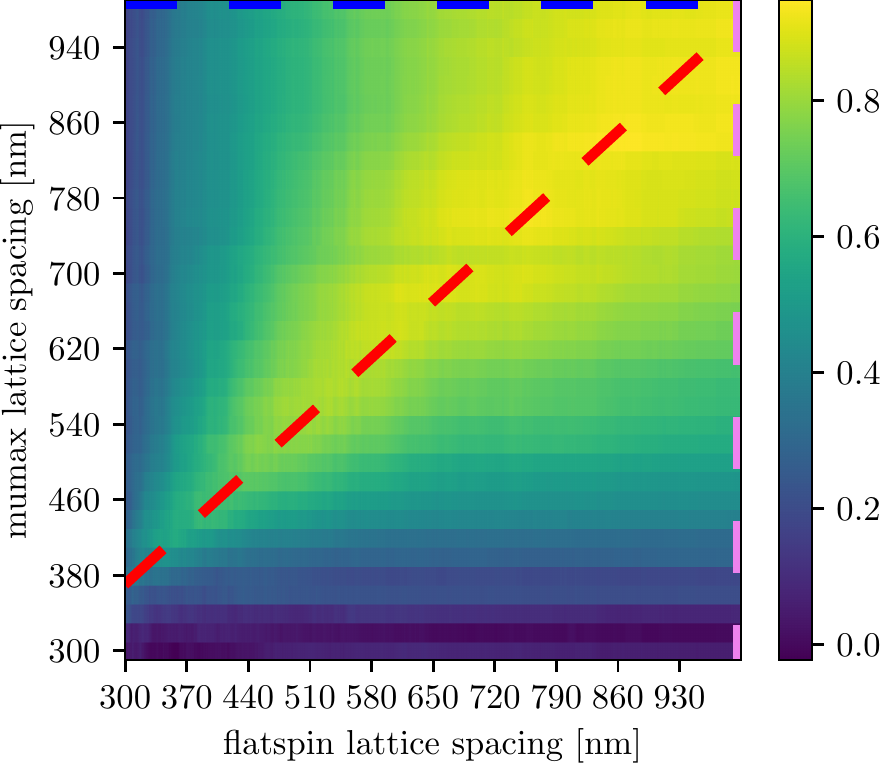}
    }
    \hspace{.2in}%
    \subfloat[\label{fig:fs_mm_spearmanb}] {
        \includegraphics[height=2.3in,width=0.4\textwidth]{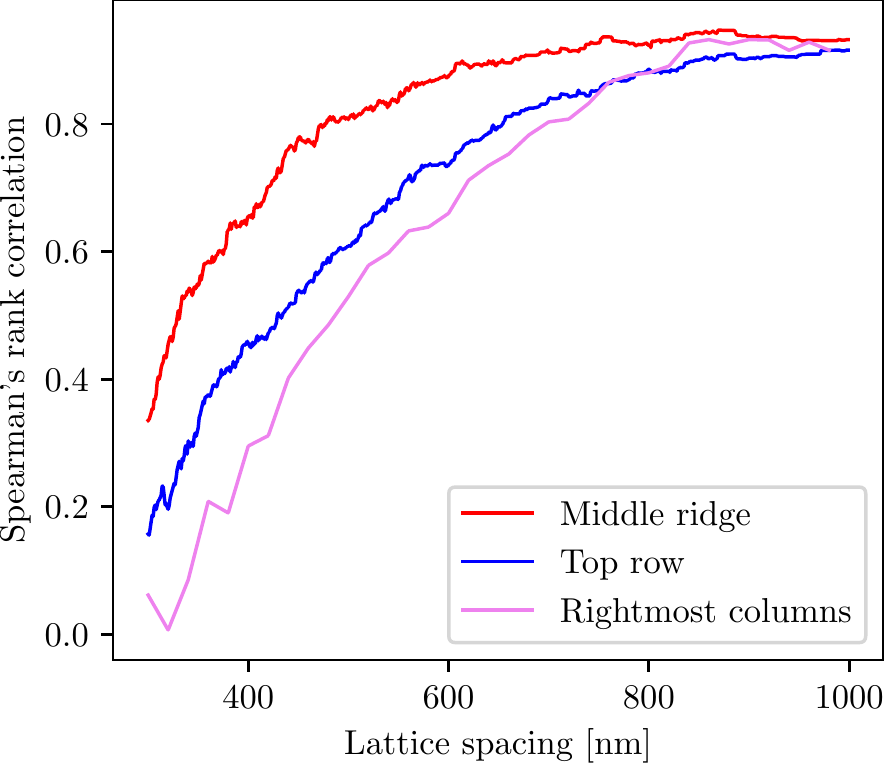}
    }
    \caption{\label{fig:fs_mm_spearman}
    (a) Spearman's rank correlation coefficients $\rho$ averaged over 32 different square ASIs, evaluated for different lattice spacings in flatspin and MuMax3.
    The red line shows the approximate maximum ridge line through the heatmap.
    (b) The red line shows the true maximum $\rho$ for the lattice spacing pairs. 
    The blue and violet lines show projections of the top row and rightmost column of (a), respectively.
    }%
\end{figure*}

\Cref{fig:li2019-theta0,fig:pw-theta0,fig:li2019-theta6,fig:pw-theta6} show hysteresis loops and array snapshots when the field is aligned with the array ($\theta=0\degree$ and $\theta=-6\degree$).
As can be seen, the results from flatspin (\cref{fig:pw-theta0,fig:pw-theta6}) are qualitatively very similar to experimental results (\cref{fig:li2019-theta0,fig:li2019-theta6}).
In all cases, the ASI undergoes reversal in a single avalanche.
Reversal begins at a small number of nucleation points close to the edge, followed by domain growth and domain wall movement perpendicular to the direction of the field.
The simulated system appears to have an anisotropy axis of $0\degree$ as opposed to $-6\degree$ observed experimentally.
Hence \cref{fig:pw-theta0} is most similar to \cref{fig:li2019-theta6} and \cref{fig:pw-theta6} is most similar to \cref{fig:li2019-theta0}.
It should be noted that the non-zero anisotropy axis found experimentally has not yet been explained.

\Cref{fig:li2019-theta30,fig:pw-theta30} show the hysteresis loops and array snapshots when the field is misaligned with the array ($\theta=30\degree$).
Again, flatspin simulations (\cref{fig:pw-theta30}) replicate key features observed experimentally (\cref{fig:li2019-theta30}).
Reversal now happens in two steps: the sublattice whose magnets have their easy axis most aligned with the field will switch first, followed later by the other sublattice.
This two-step reversal process results in an emergent rotation of the collective magnetization.
The magnetization is constrained to follow the orientation of the magnets, resulting in reversal via stripe patterns at $45\degree$.

\textcite{Li2019} report they were unable to replicate the magnetization details using a point-dipole Monte Carlo model.
One crucial difference between flatspin and their dipole model is the switching criteria.
They use the simpler criteria $\vec{h}_i \cdot \vec{m}_i < h_{k}^{(i)}$, which considers only the parallel field component and will be largely inaccurate for fields which are not aligned with the magnet's easy axis.
Indeed, using this simpler switching criteria in flatspin results in a very different reversal process and magnetization details (not shown).

\subsection{Comparison to micromagnetic single-spin switching order}

Micromagnetic simulations in MuMax3\cite{mumax3} have been shown to agree with experiment due to high simulation fidelity.
It is, therefore, of interest to study how well flatspin agrees with MuMax3 at the level of detail expressed in flatspin. 

Here we evaluate the switching strategy outlined in \Cref{sec:dynamics}, by comparing the switching orders obtained in flatspin and MuMax3, of a square ASI as it undergoes reversal by an external field.
As a similarity measure, Spearman's rank correlation coefficient $\rho$ \citep{gibbons2014nonparametric} is used, where a value of 1 indicates perfect correlation and 0 indicates no correlation between switching orders.

In the weakly coupled regime, the switching order is dictated by the coercive fields of each individual magnet.
In flatspin, the coercive field can be set directly by modifying $h_\text{k}^{(i)}$.
In MuMax3, we control the coercive field implicitly, by varying the first-order, uniaxial, magnetocrystalline anisotropy, $K_\text{U1}^{(i)}$ of each magnet.
Given a set of randomly drawn $K_\text{U1}^{(i)}$ values, the corresponding $h_\text{k}^{(i)}$ values were obtained by a linear map.

The system we considered was a $4 \times 4$ square (closed) ASI, each magnet measuring \SI{220 x 80 x 25}{\nano\meter}. 
flatspin was run with parameters $b=0.38$, $c=1$, $\beta=1.5$, and $\gamma=3.2$.
In both simulators, we applied a gradually increasing reversal field at $\theta=44\degree$. 

At a certain point the dipolar interactions begin contributing to the switching order.
To verify that flatspin still captures switching dynamics, we perform a comparision of the switching orders for all pairs of lattice spacings in both simulators.

\Cref{fig:fs_mm_spearmana} shows the correlations for each pair of lattice spacings as an average over 32 different square ASIs.
We observe a clear linear relationship between the two simulators, with higher lattice spacings exhibiting higher correlation.
The non-zero y-intercept in the heatmap indicates that, as suspected, the coupling strength is slightly underestimated by the dipole approximation employed in flatspin, in particular for lower lattice spacings.
For example, flatspin with \SI{300}{\nano\meter} lattice spacing is most similar to MuMax3 with \SI{380}{\nano\meter}.

The red line in \cref{fig:fs_mm_spearmanb} traces the ridge in the heatmap, i.e., the highest $\rho$, for each flatspin lattice spacing.
As can be seen, a near-perfect agreement between the simulators is found in the weakly coupled regime (high lattice spacing). 
As lattice spacings decrease, the effect of dipole interactions become apparent. 
Below \SI{450}{\nano\meter} the correlation drops.
Since flatspin does not account for the micromagnetic state, complete correlation is not expected.

The particular selection of $h_\text{k}^{(i)}$ values introduces an \emph{inherent} bias in the switching order.
One might expect that this bias causes the dipole interactions to have a negligible impact on the switching order, leading to an inflated correlation between flatspin and MuMax3, regardless of lattice spacing.

The violet and blue lines, (rightmost column and top row from \cref{fig:fs_mm_spearmana}), show the deviation from the inherent switching order in MuMax3 and flatspin, respectively.
Their rapid decline confirms that the switching order is not dominated by the inherent bias for highly coupled systems.
The red line clearly shows a stronger agreement between MuMax3 and flatspin (higher than the blue and violet lines), even at smaller lattice spacings.

\section{Performance}
Although the total simulation time will depend on many factors, it is of interest to measure how simulation time scales with the number of spins.
As the number of spins are increased, simulation time will be largely dominated by the calculation of the effective field, $\vec{h}_i$, acting on each of the $N$ spins in the lattice.
Computing time for $\vec{h}_\text{dip}^{(i)}$ depends on the number of neighbors around spin $i$, which is typically constant for all spins except the ones at the edges of the geometry.
For large $N$, the number of edge magnets is negligible (in the common ASI geometries).
Computing $\vec{h}_i$ for all spins will take $\mathcal{O}(N)$ time, i.e., computation time grows no faster than linear in $N$.

\begin{figure}[pt]
    \centering
    \includegraphics{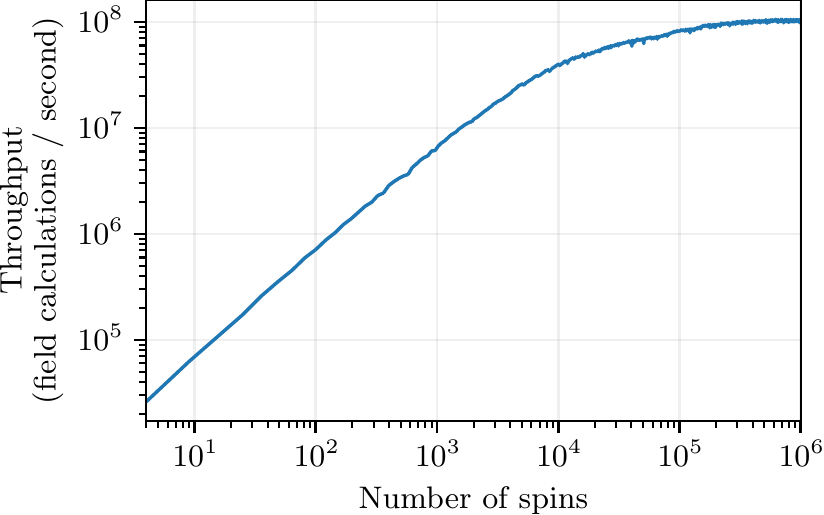}
    \caption{\label{fig:perf-throughput}%
    The throughput (number of field calculations per second) as a function of number of spins.
    Throughput is averaged over 100 simulations of each size.
    The test was performed on an NVIDIA Tesla V100 GPU with 32GB of RAM.
    Note the logarithmic scale of the axes.
    }
\end{figure}

\Cref{fig:perf-throughput} shows the throughput (number of field calculations per second) as a function of the number of spins.
Here a field calculation is defined as the computation of $\vec{h}_i$ for a single spin $i$, hence for $N$ spins there will be $N$ such field calculations.
The geometry used was square ASI (open edges) using a standard 8 spin neighborhood for calculating $\vec{h}_\text{dip}^{(i)}$.
The throughput was averaged over 100 simulations of each size.
The test was performed on an NVIDIA Tesla V100 GPU with 32GB of RAM.

At around 200 000 spins, the throughput saturates at $10^8$ field calculations per second.
On our test setup, computing $\vec{h}_i$ for one million spins takes approximately \SI{10}{\ms}.
Above 200 000 spins we are able to fully utilize the GPU resources.

To simulate the reversal of an ASI by a gradually increasing external field, at least one field calculation per spin flip is required, i.e., at least $N$ field calculations.
If the external field gradually changes with a resolution of $K$ values, the worst case will be when all spins flip during a single field value.
In this case the number of field calculations required will be $N + K - 1$ since there will be $K - 1$ field calculations which results in no spin flips.

The total simulation time depends largely on the particular experimental setup, parameters and other system characteristics.
Time will be spent on things other than field calculations, e.g., organizing and writing results to storage.
Hence the total simulation time will be longer than predicted by field calculations alone.
As an example, the simulations from \cref{sec:pinwheel} of $25 \times 25$ pinwheel ASI with 1250 magnets took approximately 6 seconds with $K=2500$, for one reversal.

\begin{figure}[t]
    \centering
    \includegraphics[width=0.4\textwidth]{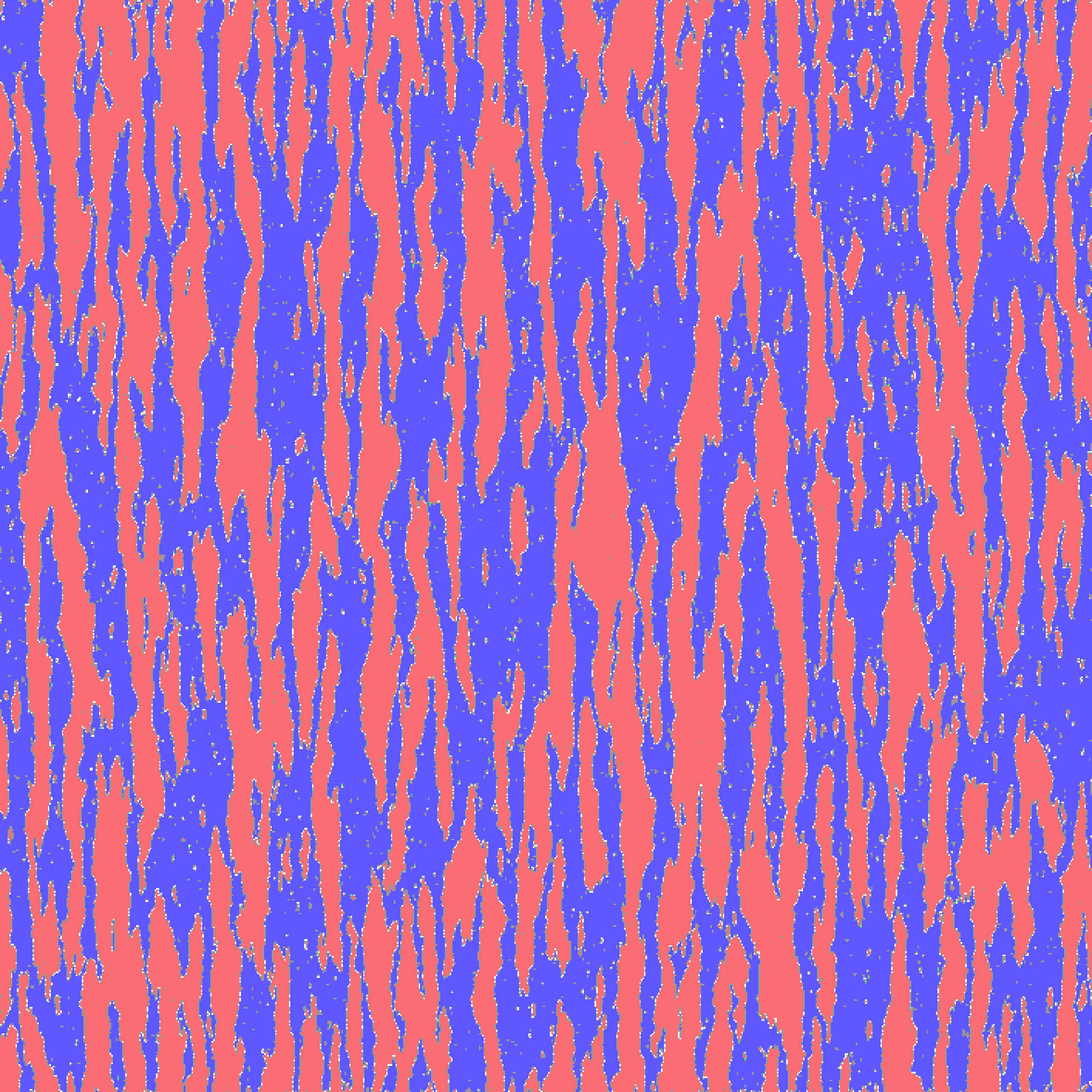}
    \caption{\label{fig:pw-1M}%
    A snapshot from flatspin simulations of $708 \times 708$ pinwheel ASI with more than one million magnets, as it undergoes reversal by an external field.
    The angle of the external field is $\theta=0\degree$.
    }
\end{figure}

\Cref{fig:pw-1M} shows a snapshot from flatspin simulations of $708 \times 708$ pinwheel ASI as it undergoes reversal by an external field.
The ability to simulate such large systems allows an experimenter to explore phenomena at much larger scales than can be directly observed experimentally.
With more than one million magnets, the simulation of array reversal took several days to complete.
A video of the full reversal is available as Supplemental Material\footnote{See Supplemental Material at [URL will be inserted by publisher] for a video of the reversal of 708x708 pinwheel ASI.}.

\section{Conclusion}
flatspin is a highly effective simulator for ASI ensemble dynamics.
At its heart lies a robust magnetic model based on dipole-dipole interactions with a switching criteria based on a generalized Stoner-Wohlfarth model.
Accompanying the model is a toolbox of useful input encoders and analysis tools. 
The model includes several common ASI geometries, and there are no inherent limits to the range of possible geometries. 

The flatspin ASI model has been verified against micromagnetic simulations and experimental results from the literature.
On a detailed level, we found good agreement between micromagnetic simulations and flatspin in terms of magnet switching order.
Emergent fine-scale patterns in kagome ASI were replicated successfully, where the formation of Dirac strings matched experimental results.
Large-scale domain sizes in square ASI were reproduced, and good agreement was found between flatspin and experimental results.
Finally, using flatspin, the experimental magnetization reversal of pinwheel ASI was reproduced for the first time in a dipole model.

Through GPU acceleration, flatspin scales to large ASI systems with millions of magnets.
High speed, parallel computation allows for a large number of ASI simulations to be executed, enabling quick exploration of parameters and novel geometries.
The flexibility and performance offered by flatspin opens for unprecedented possibilities in ASI research.

\begin{acknowledgments}
This work was funded in part by the Norwegian Research Council under the SOCRATES project, grant number 270961.
Simulations were executed on the NTNU EPIC compute cluster\cite{Epic2019}.
\end{acknowledgments}

\bibliographystyle{apalike}
\bibliography{bibliography}

\begin{thebibliography}{}

\bibitem[Anghinolfi et~al., 2015]{anghinolfi2015}
Anghinolfi, L., Luetkens, H., Perron, J., Flokstra, M.~G., Sendetskyi, O.,
  Suter, A., Prokscha, T., Derlet, P.~M., Lee, S.~L., and Heyderman, L.~J.
  (2015).
\newblock Thermodynamic phase transitions in a frustrated magnetic
  metamaterial.
\newblock {\em Nature Communications}, 6(1):8278.

\bibitem[Bar-Yam, 1997]{baryam}
Bar-Yam, Y. (1997).
\newblock {\em Dynamics of Complex Systems}.
\newblock Perseus Books, USA.

\bibitem[Budrikis, 2012]{budrikis_phd}
Budrikis, Z. (2012).
\newblock {\em Athermal dynamics of artificial spin ice: disorder, edge and
  field protocol effects}.
\newblock PhD thesis, The University of Western Australia.

\bibitem[Fan and Wu, 1969]{chungpeng-ising-neighbors}
Fan, C. and Wu, F.~Y. (1969).
\newblock Ising model with second-neighbor interaction. i. some exact results
  and an approximate solution.
\newblock {\em Phys. Rev.}, 179:560--569.

\bibitem[flatspin contributors, 2020a]{FlatspinUserManual}
flatspin contributors (2020a).
\newblock flatspin user manual.
\newblock \url{https://flatspin.readthedocs.io/en/latest/}.

\bibitem[flatspin contributors, 2020b]{FlatspinWebsite}
flatspin contributors (2020b).
\newblock flatspin website.
\newblock \url{https://flatspin.gitlab.io}.

\bibitem[Frenkel and Doefman, 1930]{frenkel-singledomain}
Frenkel, J. and Doefman, J. (1930).
\newblock Spontaneous and induced magnetisation in ferromagnetic bodies.
\newblock {\em Nature}, 126(3173):274--275.

\bibitem[Gibbons and Chakraborti, 2014]{gibbons2014nonparametric}
Gibbons, J.~D. and Chakraborti, S. (2014).
\newblock {\em Nonparametric Statistical Inference: Revised and Expanded}.
\newblock CRC press.

\bibitem[Gillette and Oshima, 1958]{Gillette1958-nanoswitching}
Gillette, P.~R. and Oshima, K. (1958).
\newblock Magnetization reversal by rotation.
\newblock {\em Journal of Applied Physics}, 29(3):529--531.

\bibitem[Heyderman and Stamps, 2013]{Heyderman2013}
Heyderman, L.~J. and Stamps, R.~L. (2013).
\newblock {Artificial ferroic systems: novel functionality from structure,
  interactions and dynamics}.
\newblock {\em Journal of Physics: Condensed Matter}, 25(36):363201.

\bibitem[Jensen et~al., 2018]{jensen2018}
Jensen, J.~H., Folven, E., and Tufte, G. (2018).
\newblock Computation in artificial spin ice.
\newblock In {\em The 2018 {Conference} on {Artificial} {Life}}, pages 15--22,
  Tokyo, Japan. MIT Press.

\bibitem[Ke et~al., 2008]{ke_energy_2008}
Ke, X., Li, J., Nisoli, C., Lammert, P.~E., McConville, W., Wang, R.~F.,
  Crespi, V.~H., and Schiffer, P. (2008).
\newblock Energy {Minimization} and ac {Demagnetization} in a {Nanomagnet}
  {Array}.
\newblock {\em Physical Review Letters}, 101(3):037205.

\bibitem[Kikuchi, 1956]{Kikuchi1956-nanoswitching}
Kikuchi, R. (1956).
\newblock On the minimum of magnetization reversal time.
\newblock {\em Journal of Applied Physics}, 27(11):1352--1357.

\bibitem[Kittel, 1946]{kittel-singledomain}
Kittel, C. (1946).
\newblock Theory of the structure of ferromagnetic domains in films and small
  particles.
\newblock {\em Phys. Rev.}, 70:965--971.

\bibitem[Levis et~al., 2013]{levis_thermal_2013}
Levis, D., Cugliandolo, L.~F., Foini, L., and Tarzia, M. (2013).
\newblock Thermal {Phase} {Transitions} in {Artificial} {Spin} {Ice}.
\newblock {\em Physical Review Letters}, 110(20):207206.

\bibitem[Li et~al., 2019]{Li2019}
Li, Y., Paterson, G.~W., Macauley, G.~M., Nascimento, F.~S., Ferguson, C.,
  Morley, S.~A., Rosamond, M.~C., Linfield, E.~H., MacLaren, D.~A.,
  Mac{\^{e}}do, R., Marrows, C.~H., McVitie, S., and Stamps, R.~L. (2019).
\newblock {Superferromagnetism and Domain-Wall Topologies in Artificial
  “Pinwheel” Spin Ice}.
\newblock {\em ACS Nano}, page acsnano.8b08884.

\bibitem[Mac{\^{e}}do et~al., 2018]{Macedo2018}
Mac{\^{e}}do, R., Macauley, G.~M., Nascimento, F.~S., and Stamps, R.~L. (2018).
\newblock {Apparent ferromagnetism in the pinwheel artificial spin ice}.
\newblock {\em Physical Review B}, 98(1):014437.

\bibitem[Mengotti et~al., 2011]{mengotti2011}
Mengotti, E., Heyderman, L.~J., Rodr{\'\i}guez, A.~F., Nolting, F., H{\"u}gli,
  R.~V., and Braun, H.-B. (2011).
\newblock Real-space observation of emergent magnetic monopoles and associated
  dirac strings in artificial kagome spin ice.
\newblock {\em Nature Physics}, 7(1):68.

\bibitem[Morris et~al., 2009]{morris_dirac_2009}
Morris, D. J.~P., Tennant, D.~A., Grigera, S.~A., Klemke, B., Castelnovo, C.,
  Moessner, R., Czternasty, C., Meissner, M., Rule, K.~C., Hoffmann, J.-U.,
  Kiefer, K., Gerischer, S., Slobinsky, D., and Perry, R.~S. (2009).
\newblock Dirac strings and magnetic monopoles in the spin ice {Dy2Ti2O7}.
\newblock {\em Science}, 326(5951):411--414.

\bibitem[Newman and Barkema, 1999]{Newman1999}
Newman, M. E.~J. and Barkema, G.~T. (1999).
\newblock {\em Monte Carlo Methods in Statistical Physics}.
\newblock Clarendon Press.

\bibitem[Note1, ]{Note1}
Note1.
\newblock See Supplemental Material at [URL will be inserted by publisher] for
  a video of the reversal of 708x708 pinwheel ASI.

\bibitem[Qi et~al., 2008]{Qi2008}
Qi, Y., Brintlinger, T., and Cumings, J. (2008).
\newblock {Direct observation of the ice rule in an artificial kagome spin
  ice}.
\newblock {\em Physical Review B}, 77(9):094418.

\bibitem[Rougemaille et~al., 2011]{rougemaille2011}
Rougemaille, N., Montaigne, F., Canals, B., Duluard, A., Lacour, D., Hehn, M.,
  Belkhou, R., Fruchart, O., El~Moussaoui, S., Bendounan, A., et~al. (2011).
\newblock Artificial kagome arrays of nanomagnets: a frozen dipolar spin ice.
\newblock {\em Physical Review Letters}, 106(5):057209.

\bibitem[Saccone et~al., 2019]{random-ising-farhan}
Saccone, M., Scholl, A., Velten, S., Dhuey, S., Hofhuis, K., Wuth, C., Huang,
  Y.-L., Chen, Z., Chopdekar, R.~V., and Farhan, A. (2019).
\newblock Towards artificial ising spin glasses: Thermal ordering in randomized
  arrays of ising-type nanomagnets.
\newblock {\em Phys. Rev. B}, 99:224403.

\bibitem[Själander et~al., 2019]{Epic2019}
Själander, M., Jahre, M., Tufte, G., and Reissmann, N. (2019).
\newblock {EPIC}: An energy-efficient, high-performance {GPGPU} computing
  research infrastructure.
\newblock arXiv:1912.05848 [cs.DC].

\bibitem[Skjærvø et~al., 2020]{skjaervo_advances_2020}
Skjærvø, S.~H., Marrows, C.~H., Stamps, R.~L., and Heyderman, L.~J. (2020).
\newblock Advances in artificial spin ice.
\newblock {\em Nature Reviews Physics}, 2(1):13--28.

\bibitem[Sklenar et~al., 2019]{sklenar2019}
Sklenar, J., Lao, Y., Albrecht, A., Watts, J.~D., Nisoli, C., Chern, G.-W., and
  Schiffer, P. (2019).
\newblock Field-induced phase coexistence in an artificial spin ice.
\newblock {\em Nature Physics}, 15(2):191--195.

\bibitem[Tanaka et~al., 2006]{Tanaka2006}
Tanaka, M., Saitoh, E., Miyajima, H., Yamaoka, T., and Iye, Y. (2006).
\newblock {Magnetic interactions in a ferromagnetic honeycomb nanoscale
  network}.
\newblock {\em Physical Review B}, 73(5):052411.

\bibitem[Tannous and Gieraltowski, 2008]{stoner-wohlfarth-Tannous_2008}
Tannous, C. and Gieraltowski, J. (2008).
\newblock The stoner{\textendash}wohlfarth model of ferromagnetism.
\newblock {\em European Journal of Physics}, 29(3):475--487.

\bibitem[Vansteenkiste et~al., 2014]{mumax3}
Vansteenkiste, A., Leliaert, J., Dvornik, M., Helsen, M., Garcia-Sanchez, F.,
  and {Van Waeyenberge}, B. (2014).
\newblock {The design and verification of MuMax3}.
\newblock {\em AIP Advances}, 4(10):107133.

\bibitem[Wang et~al., 2006]{Wang2006}
Wang, R.~F., Nisoli, C., Freitas, R.~S., Li, J., McConville, W., Cooley, B.~J.,
  Lund, M.~S., Samarth, N., Leighton, C., Crespi, V.~H., and Schiffer, P.
  (2006).
\newblock {Artificial 'spin ice' in a geometrically frustrated lattice of
  nanoscale ferromagnetic islands}.
\newblock {\em Nature}, 439(7074):303--306.

\bibitem[Zhang et~al., 2013]{Zhang2013}
Zhang, S., Gilbert, I., Nisoli, C., Chern, G.~W., Erickson, M.~J., O'Brien, L.,
  Leighton, C., Lammert, P.~E., Crespi, V.~H., and Schiffer, P. (2013).
\newblock {Crystallites of magnetic charges in artificial spin ice}.
\newblock {\em Nature}, 500(7464):553--557.

\end{thebibliography}
\end{document}